\documentclass[sigconf]{acmart}

\usepackage{enumitem}
\usepackage{color}
\usepackage{float}
\usepackage{stfloats}
\usepackage{soul}
\usepackage{tabularx}
\usepackage{multirow}
\usepackage{booktabs}
\usepackage{hyperref}
\usepackage{xcolor}
\usepackage{makecell}
\usepackage{graphicx}
\usepackage{amsmath}
\usepackage{listings}
\usepackage{fvextra}
\usepackage{subcaption}
\usepackage{array}
\usepackage{svg}
\usepackage{tikz}
\usepackage{framed}
\usepackage{fancyvrb}
\usepackage{ifthen}

\newcommand{\sysname}{\textsc{Evalet}}
\newcommand{\approach}{\textit{functional fragmentation}}
\newcommand{\Approach}{\textit{Functional fragmentation}}
\newcommand{\treatment}{\texttt{Fragmented}}
\newcommand{\control}{\texttt{Holistic}}
\newcommand{\stats}[7]{($\text{\treatment{}}=#1\pm#2$, $\text{\control{}}=#3\pm#4$, $#5=#6$, $p#7$)}
\newcommand{\myquote}[1]{\textit{``#1''}}
\definecolor{syscolor}{HTML}{3788F7}
\newcommand{\systemText}[1]{\textcolor{syscolor}{\texttt{#1}}}
\newenvironment{block}%
  {\list{}{\leftmargin=0.2in\rightmargin=0in}\item[]\color{blockcolor}}%
  {\endlist}
  
\newcommand{\criterion}[1]{
  \tikz[baseline=(text.base)] {
    \node (text) [
      fill=gray!10,       
      rounded corners=2pt, 
      inner sep=2pt,      
      anchor=base         
    ] {\texttt{#1}};
  }%
}

\definecolor{darkgreen}{HTML}{006400}
\definecolor{darkorange}{HTML}{CE8E1A}
\definecolor{teal}{HTML}{00BAAB}
\definecolor{brown}{HTML}{800000}
\definecolor{purple}{HTML}{BF4DDA}
\definecolor{navy}{HTML}{002455}

\newcommand*{\inlineimage}[1]{
    \raisebox{-.3\baselineskip}{
        \includegraphics[
        height=\baselineskip,
        width=\baselineskip,
        keepaspectratio,
        ]{#1}
    }
}

\definecolor{syscolor}{HTML}{3788F7}
\definecolor{blockcolor}{HTML}{555555}

\begin{document}

\copyrightyear{2026}
\acmYear{2026}
\setcopyright{cc}
\setcctype{by}
\acmConference[CHI '26]{Proceedings of the 2026 CHI Conference on Human Factors in Computing Systems}{April 13--17, 2026}{Barcelona, Spain}
\acmBooktitle{Proceedings of the 2026 CHI Conference on Human Factors in Computing Systems (CHI '26), April 13--17, 2026, Barcelona, Spain}
\acmPrice{}
\acmDOI{10.1145/3772318.3790285}
\acmISBN{979-8-4007-2278-3/2026/04}

\title[\sysname{}: Evaluating Large Language Models through Functional Fragmentation]{\sysname{}: Evaluating Large Language Models through\\Functional Fragmentation}

\author{Tae Soo Kim}
\email{taesoo.kim@kaist.ac.kr}
\authornote{Both authors contributed equally to this research.}
\affiliation{%
  \institution{School of Computing, KAIST}
  \city{Daejeon}
  \country{Republic of Korea}
}

\author{Heechan Lee}
\email{heechan@kaist.ac.kr}
\authornotemark[1]
\affiliation{%
  \institution{School of Computing, KAIST}
  \city{Daejeon}
  \country{Republic of Korea}
}

\author{Yoonjoo Lee}
\email{lyoonjoo@umich.edu}
\affiliation{%
  \institution{Computer Science and Engineering, University of Michigan}
  \city{Ann Arbor, MI}
  \country{USA}
}

\author{Joseph Seering}
\email{seering@kaist.ac.kr}
\affiliation{%
  \institution{School of Computing, KAIST}
  \city{Daejeon}
  \country{Republic of Korea}
}

\author{Juho Kim}
\email{juhokim@kaist.ac.kr}
\affiliation{%
  \institution{School of Computing, KAIST}
  \city{Daejeon}
  \country{Republic of Korea}
}
\email{juho@skillbench.com}
\affiliation{
  \institution{SkillBench}
  \city{Santa Barbara, CA}
  \country{USA}
}

\renewcommand{\shortauthors}{Tae Soo Kim*, Heechan Lee*, Yoonjoo Lee, Joseph Seering, and Juho Kim}

\begin{abstract}
    Practitioners increasingly rely on Large Language Models (LLMs) to evaluate generative AI outputs through "LLM-as-a-Judge" approaches. However, these methods produce holistic scores that obscure which specific elements influenced the assessments. We propose \approach{}, a method that dissects each output into key fragments and interprets the rhetoric functions that each fragment serves relative to evaluation criteria—surfacing the elements of interest and revealing how they fulfill or hinder user goals. We instantiate this approach in \sysname{}, an interactive system that visualizes fragment-level functions across many outputs to support inspection, rating, and comparison of evaluations. A user study (N=10) found that, while practitioners struggled to validate holistic scores, our approach helped them identify 48\% more evaluation misalignments. This helped them calibrate trust in LLM evaluations and rely on them to find more actionable issues in model outputs. Our work shifts LLM evaluation from quantitative scores toward qualitative, fine-grained analysis of model behavior.
\end{abstract}

% CCS


% KEYWORDS

% TEASER FIGURE
\begin{teaserfigure}
  \centering
  \includegraphics[width=0.98\textwidth]{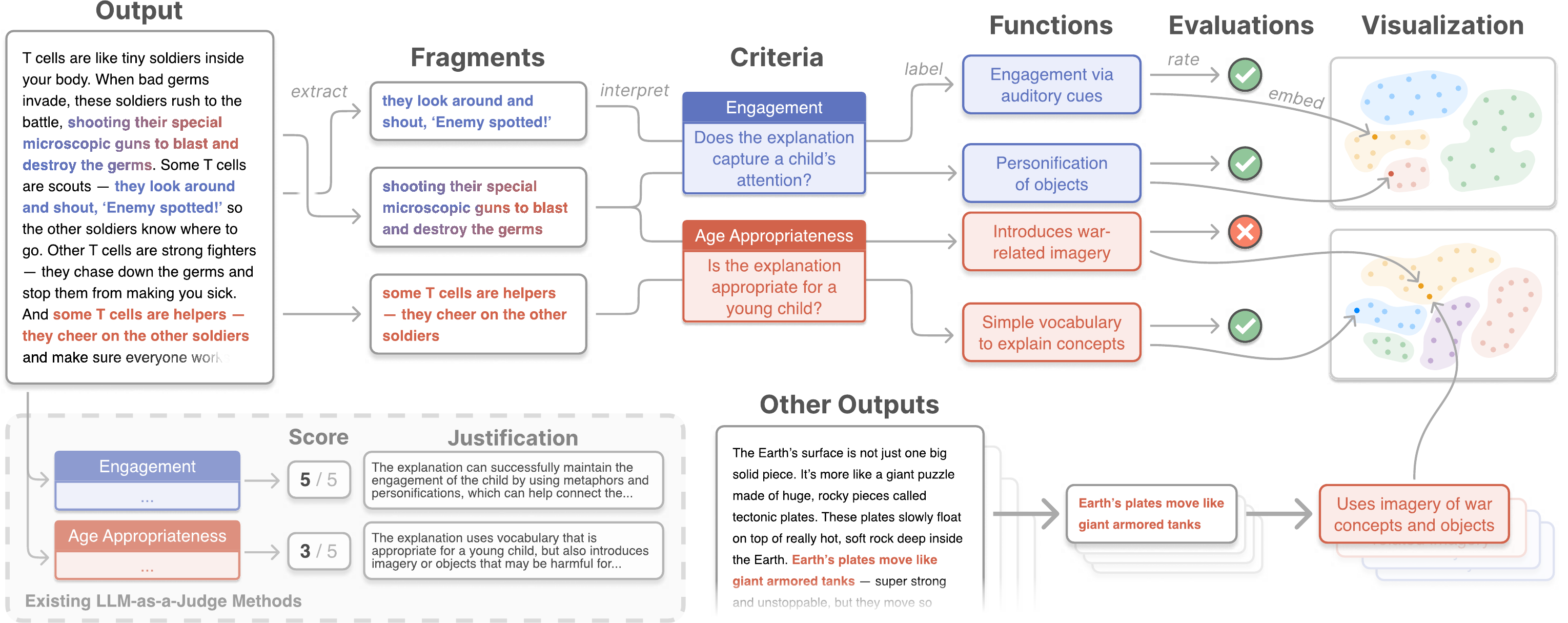}
  \caption{Illustration of the \approach{} approach supported by \sysname{}. Unlike prior approaches that evaluate LLM outputs by producing holistic numeric scores and justifications, \sysname{} extracts significant text fragments from each output. Then, the system interprets and labels the \textit{function} that each fragment plays in terms of the criterion, and rates whether the function satisfies or fails to meet the criterion. Finally, \sysname{} embeds fragment-level functions across various outputs into the same space to support interpretation and validation at scale.}
  \Description{This figure illustrates how Evalet evaluates LLM outputs using functional fragmentation. An example output on the left is broken into highlighted text fragments, which are linked to evaluation criteria such as Engagement and Age Appropriateness. The system identifies each fragment’s function, for example, Engagement via auditory cues like “they look around and shout,” and, from Age Appropriateness perspective, introducing war-related imagery through phrases like “shooting their special microscopic guns.” Evalet rates whether each function satisfies the criterion. Functions from multiple outputs are embedded into a visualization space, enabling interpretation and validation at scale.}
  \label{fig:teaser}
\end{teaserfigure}

\maketitle

\section{Introduction}

Large Language Models (LLMs) have enabled practitioners (e.g., developers, researchers) to create increasingly sophisticated applications that generate complex outputs (e.g., stories~\cite{chung2022talebrush, yuan2022wordcraft}, research papers~\cite{lu2024ai, starace2025paperbench}, and reasoning traces~\cite{jaech2024openai, snell2024scaling}).
Deploying these models safely requires rigorous verification that the outputs \textit{align}~\cite{shen2024towards} with practitioners' intended goals.
Evaluation is frequently manual as the applications are novel---lacking established benchmarks---and involve subjective aspects that require qualitative judgments~\cite{kim2024evallm}, like how insightful or harmful the application's outputs are.
Identifying systemic and recurring issues requires reviewing hundreds of outputs, but the burden of manual inspection often leads practitioners to overgeneralize from small samples~\cite{kim2024evallm, arawjo2024chainforge, szymanski2024comparing, shankar2024validates}.
To address this, practitioners have begun employing LLM-based evaluators (i.e., \textit{LLM-as-a-Judge}~\cite{zheng2023judging}), where one LLM evaluates another's outputs.
By describing multiple criteria (e.g., \criterion{Insightfulness}, \criterion{Harmlessness}) in natural language, practitioners can assess the alignment of the model outputs with their various goals~\cite{kim2023prometheus, zhong2022unieval, fu2023gptscore, liu2023geval}.

Current LLM-as-a-Judge approaches use \textit{holistic scores}, where an entire output is summarized into numeric ratings (e.g., 3 out of 5) for each criterion.
Holistic scores help practitioners quickly assess overall performance~\cite{kim2024evallm} but obscure the specific elements in the outputs that led to these assessments.
For example, in Figure~\ref{fig:teaser}, an LLM explaining \textit{``T cells''} to a young child received a moderate score for the criterion \criterion{Age Appropriateness}. 
To understand this rating, users must manually review the output to notice that while it uses simple vocabulary, it also employs potentially harmful war imagery. 
This manual process provides necessary insights but undermines the automation benefits.
While some LLM evaluators provide brief justifications, practitioners must still map the justification to the specific fragments in the output~\cite{kim2024evallm}.
This lack of detail or granularity becomes more critical at scale.
When multiple outputs receive identical scores, practitioners have to read the justifications for each output's evaluation to determine whether they share the same issues or different ones.
Ultimately, the opaqueness of holistic scores inhibits practitioners from identifying systemic failure patterns in the outputs that require urgent attention~\cite{cabrera2023did, ribeiro2020beyond}, and validating the accuracy and consistency of the LLM evaluator's judgments~\cite{gebreegziabher2025metricmate}.

To address these challenges, we propose \approach{} (Fig.~\ref{fig:teaser}): a novel LLM-based evaluation method that dissects each output into key \textbf{fragments} and interprets the \textbf{functions} of each fragment, where each fragment may serve multiple functions.
With \textit{functions}, we refer to \textbf{the rhetorical roles or purposes that text fragments serve that are relevant to a given evaluation criterion.}
In Figure~\ref{fig:teaser}, the fragment describing T cells as \textit{``shooting their special microscopic guns''} serves the function of \textit{``personification''} for the criterion \criterion{Engagement}, but also \textit{``war-related imagery''} for \criterion{Age Appropriateness}.

We propose that disentangling outputs into fragment-level functions supports new interaction affordances for \textbf{inspecting}, \textbf{rating}, and \textbf{comparing} outputs.
We instantiate \approach{} and these affordances in \sysname{}, an interactive system for analyzing LLM outputs based on fragment-level functions surfaced for criteria defined by the user.
For \textbf{inspection}, \sysname{} summarizes each output into lists of the surfaced functions per criterion---allowing users to jump directly to elements of interest and verify their interpretations, instead of manually scanning the whole output and mapping justifications to the output.
For \textbf{rating}, \sysname{} individually assesses each function's alignment with the criterion to provide more interpretable scores based on the proportion of aligned to misaligned functions---rather than opaque numeric scores.
Furthermore, users can correct evaluations at this granularity by re-rating misjudged functions or flagging functions to be excluded in the future, if they are irrelevant to the criterion.
For \textbf{comparison}, \sysname{} pools fragments from all outputs, and then projects and clusters them in a two-dimensional space based on the similarity of their functions, rather than their lexical content.
Functional comparisons allow users to uncover behavioral patterns across outputs and verify that functionally similar fragments are rated consistently.
For example, in Figure~\ref{fig:teaser}, fragments with different wording (e.g., \textit{``shooting [...] microscopic guns''} and \textit{``move like giant armored tanks''}) are grouped as they serve functions related to war themes. 
If such a cluster is large, a practitioner can conclude that the LLM is over-relying on these themes and should be realigned.

To understand how users analyze models and validate evaluations with \sysname{}, we conducted a within-subjects study with practitioners (N=10) comparing \sysname{} against a baseline that only provides holistic scores and justifications, like existing LLM-based evaluations.
Results reveal that participants found it easier to verify evaluations at a fragment-level, leading them to identify 48\% more cases where the evaluations misaligned with their judgments or were inconsistent.
Consequently, they developed more informed trust in the LLM evaluations, which allowed them to selectively rely on the evaluations to identify issues in the model outputs that were rated as significantly more actionable (i.e., higher self-confidence in acting on and resolving these issues).
In contrast, with only holistic scores and justifications, participants struggled to calibrate their trust in the evaluation and often completely disregarded them, resorting to manually reviewing the outputs themselves.
In an open-ended exploration session, participants noted how \approach{} supported a process resembling \textit{inductive coding}: given a broad theme (i.e., the criterion), the system surfaced previously unconsidered codes (i.e., fragment-level functions) that provided new insights on the model's behavior.
Overall, our work proposes that \approach{} can shift LLM evaluation from focusing on opaque and quantitative scores to a more qualitative, actionable, and fine-grained analysis of model behavior.
\section{Related Work}

This work aims to support interactive evaluation of LLMs through sensemaking of model outputs and evaluations at scale.
To this end, we review literature in (1) interactive examination of machine learning (ML) models, (2) interactive testing and evaluation of LLMs, and (3) sensemaking of text at scale.

\subsection{Interactive Examination of Machine Learning}

Traditionally, Machine Learning (ML) models are examined and understood by evaluating on benchmarks with automated metrics that aggregate performance into a single statistic or score.
However, this provides limited understanding into how the model behaves, what its flaws are, and what are the areas for improvement~\cite{ribeiro2020beyond, zeng2025evaltree, murahari2023qualeval}.
Instead of relying on aggregated metrics, prior work has introduced systems that support more interactive and fine-grained evaluation of models.
For example, \textit{Polyjuice}~\cite{wu2021polyjuice} and \textit{AdaTest}~\cite{ribeiro2022adaptive} allow practitioners to iteratively evaluate models by creating challenging input data and testing how the models behave on these cases.
Furthermore, researchers have proposed various tools~\cite{cabrera2023zeno, wexler2020whatif, wu2019errudite, robertson2023angler, sivaraman2025divisi} that help practitioners to unpack evaluations by identifying \textit{slices} or subsets of data, and testing models on these to identify specific flaws or limitations.
Complementing these evaluation approaches, a rich body of work in explainable AI (XAI)~\cite{liao2021human} has also explored how to support understanding of models through more fine-grained analysis of behaviors~\cite{lai2022human}. 
For instance, the foundational methods LIME~\cite{ribeiro2016should} and SHAP~\cite{lundberg2017unified} explain model predictions by inferring the effect of individual features.
Frameworks such as the XAI Question Bank~\cite{liao2020questioning} organizes users' understanding and explainability needs into fine-grained questions that explore diverse aspects of models (e.g., inputs, outputs, performance).
Collectively, this work highlights the importance and need for more granular analysis of model performance.
Our work extends this to the evaluation of LLMs: instead of simply slicing datasets, we slice the data points themselves into fragment-level functions to provide a more fine-grained understanding of LLM outputs.

\subsection{Interactive Testing and Evaluation of LLMs}

The general-purpose capabilities of LLMs have enabled novel AI applications but also made it harder to verify that they behave as intended.
As these models are applied to new tasks and contexts, there are no benchmarks or metrics to automate evaluation~\cite{kim2024evallm} and, due to their near infinite input-output space, models have to be tested with numerous and diverse samples~\cite{zamfirescu2023herding, liu2023what}.
To help practitioners, researchers have proposed novel tools that support interactive testing and experimentation on these models by decomposing tasks into chains of sub-tasks~\cite{wu2022aichains}, composing diverse LLM pipelines in parallel~\cite{arawjo2024chainforge, zhang2024chainbuddy}, or creating diverse variations of test inputs~\cite{strobelt2023promptide, wu2023scattershot, mishra2023promptaid, kim2023cells}.
More recently, the success of \textit{LLM-as-a-Judge}~\cite{zheng2023judging} (i.e., LLMs evaluating other LLMs) has led to several systems~\cite{kim2024evallm, kahng2024llm, ashktorab2024aligning, shankar2024validates} that employ LLM-based evaluators to support interactive evaluation on diverse criteria---offering a multi-dimensional view of model performance.
However, as these only provide holistic scores and overall justifications, practitioners must manually review the outputs and justifications to identify specific strengths and weaknesses, and validate the evaluations~\cite{kim2024evallm, gebreegziabher2025metricmate}---requiring effort that is impractical at scale.
To address the limitations of holistic scoring, prior work has proposed more fine-grained evaluation methods. 
For example, Nenkova and Passonneau~\cite{nenkova2004evaluating} and FactScore~\cite{min2023factscore} decompose text into units, which are then assessed individually, while Scarecrow~\cite{dou2021gpt} and BooookScore ~\cite{chang2023booookscore} assess outputs by annotating spans on predefined error categories~\cite{dou2021gpt, chang2023booookscore}.
Building on this, our approach employs LLMs to decompose outputs into fragments, annotate their function in an emergent manner, and cluster these across outputs---surfacing common strengths and weaknesses, and facilitating verification of evaluation consistency.

\subsection{Sensemaking of Text at Scale}

Theories on sensemaking posit that people make sense of large information spaces through multiple cognitive processes such as iteratively foraging and organizing information---\textit{the notional model}~\cite{pirolli2005sensemaking}---and comparing information to identify patterns---\textit{structure-mapping theory}~\cite{gentner1983structure}.
Given their significant cognitive demands, prior work has proposed systems that support these processes for textual data: facilitating structuring and organization of collected information~\cite{palani2022interweave, rachatasumrit2021forsense}, generating summaries or topic models~\cite{lam2024concept, yatani2011review, palani2021conotate, glassman2015overcode}, and visualizing corpora using spatial embeddings or structural patterns~\cite{kim2016topiclens, wang2023wizmap, gu2025abstractexplorer}.
Recent work extends these ideas to sensemaking over LLM outputs~\cite{suh2023sensecape, jiang2023graphologue}.
For example, \textit{Luminate}~\cite{suh2024luminate} guides LLMs to generate outputs along key dimensions and then visualizes the outputs on these dimensions, helping writers explore the generation space.
Gero et al.~\cite{gero2024supporting} explored various designs and algorithms (e.g., unique words, exact matches) to support comparison of LLM outputs and help users form mental models of LLM behavior.
Most similar to our work, \textit{Policy Projector}~\cite{lam2024ai} \textit{``maps''} LLM input-output pairs into a 2D space to help users explore common groups of outputs, classify these groups, and define policies on the model's behaviors using these classified groups.
Building on these approaches, we propose a novel approach for \textit{multi-dimensional} and \textit{granular} sensemaking of LLM outputs. 
Instead of visualizing entire outputs, we extract fragment-level functions from multiple outputs for each criterion and then visualize the space of functions for each criterion---supporting exploration of fine-grained model behaviors within dimensions of interest.
\section{Functional Fragmentation: An Evaluation Approach}

To evaluate the alignment of an LLM, practitioners must not only quantify quality through numeric scores but also qualitatively understand how this models' outputs are composed and their characteristics~\cite{ribeiro2020beyond, dunlap2024vibecheck, gero2024supporting}.
While existing LLM-based evaluation methods can assess outputs across various criteria~\cite{zheng2023judging, kim2024evallm, ye2023flask}, they only provide holistic judgments (i.e., overall scores, justification) for each dimension.
Thus, practitioners must manually inspect outputs to identify the specific elements that satisfy or violate their goals.

To address this, we introduce \approach{}, an LLM-based evaluation method that \emph{decomposes} model outputs into criterion-relevant \emph{fragments} and then infers each fragment’s \emph{function}---i.e., the role or effect it serves that influences the output's fulfillment of that criterion. 
Our approach draws inspiration from inductive coding~\cite{thomas2006general} (i.e., interpreting raw data into codes based on higher-level themes) and rubric design~\cite{perlman2003performance} (i.e., inspecting artifacts to define quality aspects to review).
In this section, we outline the novel affordances that are supported by this approach for \textbf{inspecting}, \textbf{rating}, and \textbf{comparison} of LLM outputs.

% ------------------------------------------

\subsection{Inspect}
\label{framework:inspect}

\paragraph{Fragment-Level}
To make sense of existing LLM-based evaluations, practitioners must manually review outputs, connect them with the evaluator's justifications, and interpret each fragment's significance in terms of the criteria~\cite{kim2024evallm}.
Our approach automatically extracts criteria-relevant fragments and interprets their functions with respect to each criterion---directly presenting practitioners with qualitative interpretations to inspect and verify.
As one fragment can serve different functions under different criteria, our approach allows practitioners to examine the same content from multiple perspectives and identify trade-offs.
Given that criteria are often subjective, our approach can also uncover meaningful functions that the practitioner may have not previously considered---similar to \textit{inductive coding}~\cite{thomas2006general}.
Conversely, if the LLM evaluator extracts functions that the practitioner considers irrelevant to the criterion, practitioners can directly flag these to be ignored in future evaluations.

\paragraph{Output-Level}
Beyond identifying each fragment-level function in an output, practitioners may also need to inspect how these functions appear together in the output.
For example, when evaluating \criterion{Tension} in LLM-generated horror stories, a practitioner may need to understand how the LLM uses various functions to gradually build tension in the story.
Traditionally, this would require the practitioner to read the whole story, but not all of the content may be directly related to that criterion.
With \approach{}, each output can be summarized into a list of functions related to a given criterion, allowing practitioners to easily inspect each output by focusing only on the aspects of interest.

% ------------------------------------------

\subsection{Rate}
\label{framework:rate}

\paragraph{Fragment-Level}
By disentangling outputs into fragment-level functions, each individual function's alignment with a criterion can be rated independently. 
More fine-grained evaluations can support interpretability by clearly highlighting the specific aspects of an output that are aligned or misaligned with a criterion.
Instead of correcting the LLM evaluator by editing criteria descriptions, practitioners can directly re-rate specific functions to control future evaluations---similar to how educators develop rubrics by assessing examples of student work~\cite{perlman2003performance}. 
Beyond supporting practitioners, LLM evaluators are also more consistent when performing more fine-grained evaluations through checklists~\cite{saad2024lmunit, lin2024wildbench} or rubrics~\cite{kim2023prometheus, ye2023flask, kim2024biggen}.
Unlike these approaches, which rely on predefining these checklists and rubrics, our fragment-level functions are \textit{emergent}, identified dynamically based on the outputs and criteria.

\paragraph{Output-Level}
Instead of providing uninterpretable and opaque scores (e.g., 2 out of 5) for each model output, our approach enables us to rate each output based on its proportion of aligned and misaligned fragment-level functions (e.g., 75\% of surfaced functions are aligned)\footnote{For simplicity, we opt for equal weighting of each function. As discussed in Limitations, future work can explore automatic or manual approaches for weighting the significance of each function.}.
This provides a more interpretable signal of \textit{how much} misalignment there is in an output and why---allowing practitioners to understand what are the specific errors that need to be corrected~\cite{ribeiro2020beyond}.

% ------------------------------------------

\subsection{Compare}
\label{framework:compare}

\paragraph{Fragment-Level}
Comparing fragments across outputs can reveal common model behaviors, but directly comparing raw text is difficult because fragments may differ lexically or semantically even when they serve the similar function.
For example, when we evaluate an essay-writing LLM on \criterion{Logical Coherence}, these two sentences serve the same function as cohesive devices for a conclusion despite their wording differences: \textit{``In conclusion, the trend is clear.''}, and \textit{``To sum up, it supports our view.''}
By labeling each fragment's functions, our approach allows for comparison and grouping of fragments not based on their lexical similarity, but by their functional similarity---allowing practitioners to distill high-level insights and patterns.

\paragraph{Output-Level}
By considering each output as a list of its fragment-level functions, we can also compare outputs based on whether they share a function or set of functions.
For example, practitioners could group and filter outputs based on the inclusion of a specific function of interest and even calculate the distribution of outputs that contain certain function patterns---supporting the common practice of slicing data into subsets of interest in ML evaluation~\cite{cabrera2023zeno, wu2019errudite, sivaraman2025divisi}.
Beyond comparing outputs from a single LLM, practitioners could qualitatively compare the behaviors of different LLMs by comparing the distributions of specific functions in each model's outputs.
\begin{figure*}[!b]
    \centering
    \includegraphics[width=1.00\textwidth]{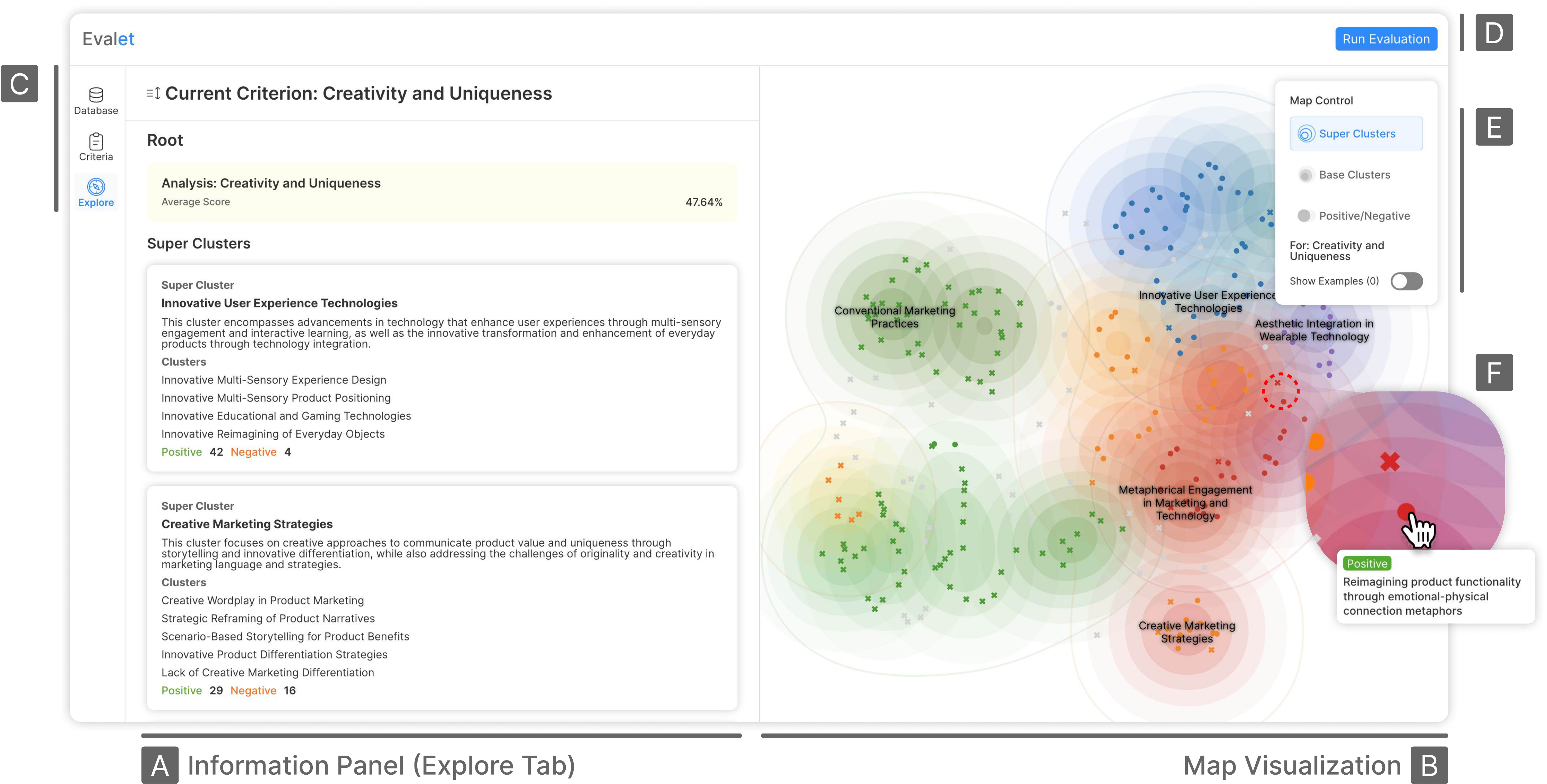}
    \caption{\sysname{} consists of two main components: (A) Information Panel and (B) Map Visualization. In the Information Panel, users can use the Tab Navigator (C) to switch between managing their input-output dataset, defining their criteria set, and viewing evaluation details. Users can initiate evaluations by clicking on \systemText{Run Evaluation} (D). The Map Visualization helps users explore all fragment-level functions across all outputs, where they can toggle what information is displayed using the Map Controls (E). Each fragment-level function is shown as a dot if rated positive or a cross if negative, and users can hover over these to see the function description (F).}
    \Description{This figure shows Evalet’s main interface, which consists of two coordinated views: an Information Panel on the left and a Map Visualization on the right. The Explore Tab in the Information Panel displays evaluation results grouped by semantic clusters for a selected criterion. Users can switch between tabs through the Tab Navigator, initiate evaluation by the Run Evaluation button, and adjust what is shown on the map via Map Controls. The Map Visualization presents each function as a point (positive) or cross (negative), clustered by semantic similarity. Hovering over a point reveals its function.}
    \label{fig:system_overview}
\end{figure*}

\section{\sysname{}: Evaluation of LLM Outputs based on Fragment-Level Functions}

To instantiate the concept of \approach{}, we present \sysname{}, an interactive system that enables users to \textbf{inspect}, \textbf{rate}, and \textbf{compare} LLM outputs at both the fragment-level and output-level.
Through an LLM-based evaluator, \sysname{} automatically disentangles outputs into fragment-level functions based on user-defined criteria, rates the alignment of each function, and visualizes the evaluations to support exploration and verification of evaluations.
\sysname{} consists of the following components:
\begin{itemize}
    \item \textbf{Input-Output Dataset}: Pairs of \textit{inputs} given to the user's LLM or LLM-based application, and pre-generated \textit{outputs} by the LLM. The user uploads this dataset to the system.
    \item \textbf{Evaluation Criteria}: Each \textit{criterion} is defined by a name and a description in natural language.
    \item \textbf{Fragment-Level Functions}: \sysname{} extracts criterion-relevant fragments and interprets the functional role or effect of each fragment to assign a short \textit{function} label. Each function receives a ``positive'' or ``negative'' rating based on its alignment with the criterion\footnote{A single fragment can be interpreted to serve different functions for different criteria, where one such function aligns with its respective criterion while the other misaligns with its criterion. As a result, we rate each \textit{function} rather than each \textit{fragment}.}.
    \item \textbf{Fragment-Level Justifications}: \sysname{} provides the LLM-based evaluator's \textit{justification} or reasoning for the rating of each fragment-level function.
    \item \textbf{Holistic Score and Justification}: For each output and criterion, \sysname{} provides a \textit{holistic score}---ratio of positive to total fragment-level functions---and a \textit{holistic justification}, a paragraph summarizing all fragment-level justifications to provide a reasoning on the overall quality of the output.
    \item \textbf{Base Clusters}: To support comparison and identification of common patterns between outputs, \sysname{} groups similar functions from different outputs into \textit{base clusters} for each criterion. Each cluster is represented by a name and a description.
    \item \textbf{Super Clusters}: Furthermore, \sysname{} also groups similar base clusters into \textit{super clusters} to provide high-level overviews of the potentially vast landscape of functions.
\end{itemize}

\subsection{Interface Walkthrough}
The user interface of \sysname{} has two main components: (1) \textbf{\textit{the Information Panel}} on the left (Fig.~\ref{fig:system_overview}A) and (2) \textbf{\textit{the Map Visualization}} on the right (Fig.~\ref{fig:system_overview}B).
The \textit{Information Panel} presents details about the LLM outputs, evaluation results, fragment-level functions, and clusters.
The \textit{Map Visualization} allows users to explore fragment-level functions and clusters in a two-dimensional space.
These views are synchronized, where information in one component is highlighted if the user interacts with relevant information in the other.
We illustrate system interactions using an example scenario where a developer, Robin, implements an LLM-based application that generates short advertisement posts from product descriptions.

\subsubsection{Initializing Data and Criteria Set}
When the user first enters the system, they upload their input-output dataset in the \textit{Database Tab} \inlineimage{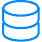} in the Information Panel.
Then, they can define their criteria in the \textit{Criteria Tab} \inlineimage{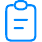} and click on \systemText{``Run Evaluation''} (Fig.~\ref{fig:system_overview}D) to evaluate the outputs on the criteria.

\begin{block}
  To test her application, Robin uploads a dataset of 100 product descriptions and the advertisement generated for each product into \sysname{}.
  In the Criteria Tab, she defines two criteria---\criterion{Creativity and Uniqueness} and \criterion{Emotional Effect}---to evaluate whether the generated advertisements are creative and engaging.
\end{block}

\begin{figure*}[!t]
    \centering
    \includegraphics[width=1.00\textwidth]{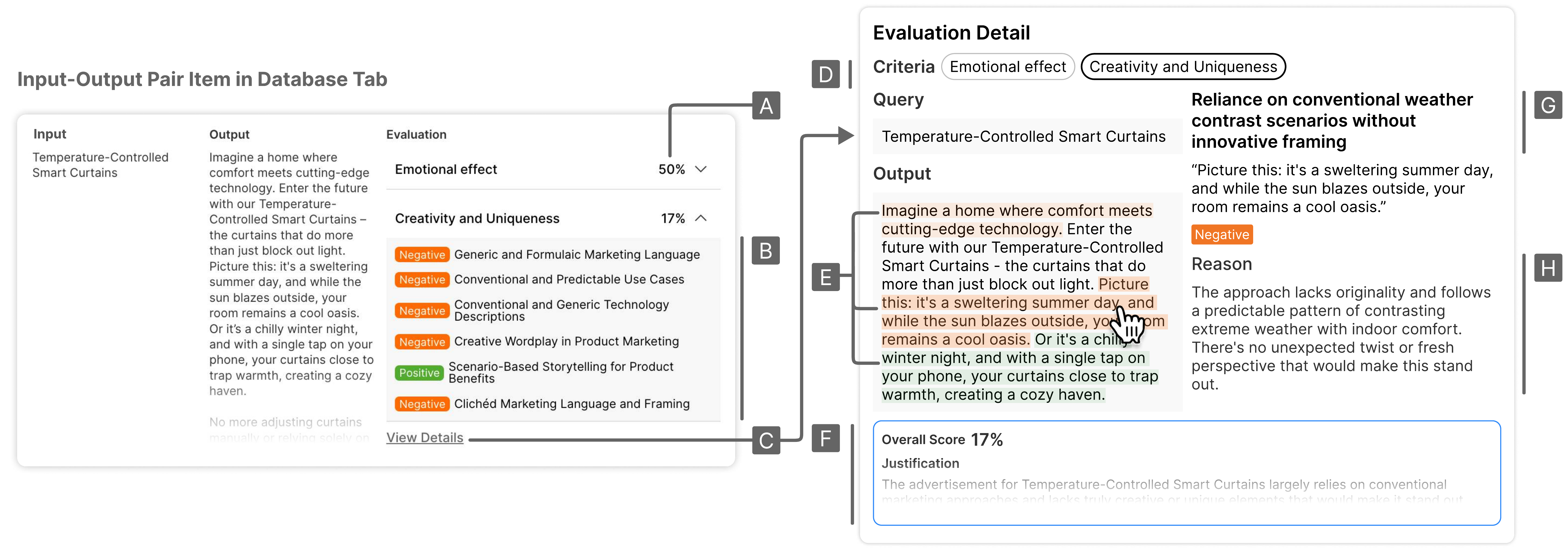}
    \caption{In the Database Tab, users can view their dataset of input-output pairs. Each item consists of the input, the output, and an evaluation summary. This summary presents the output's holistic score on each criterion (A) and its list of fragment-level functions (B). Users can see more details by clicking on \systemText{View Details} (C). On the details page, the user selects a criterion to view the relevant evaluations (D). Assessed fragments from the output are highlighted in green if positive and orange if negative (E). The bottom of the interface displays the holistic score and justification provided by the LLM (F). By clicking on each fragment, users can view the corresponding function description (G) and the evaluator's reasoning in detail (H).}
    \Description{This figure shows how users interact with Evalet’s Database Tab to inspect evaluations of individual input-output pairs. Each item displays the input, generated output, and an evaluation summary, including holistic criterion scores and including functions. Clicking "View Details" opens the Evaluation Detail view where users can select a criterion to highlight relevant fragments in the output—green for positive, orange for negative. The bottom of the panel shows the holistic score and justification. Clicking a highlighted fragment reveals its function label and the reasoning behind its evaluation.}
    \label{fig:system_database}
\end{figure*}

\subsubsection{Inspecting Evaluation Results}

After the evaluation completes, the user can navigate to the \textit{Database Tab} to skim through each input-output pair and gauge overall quality through the holistic scores on each criterion (Fig.~\ref{fig:system_database}A).
For more detail, the user can open the evaluation summary for a criterion (Fig.~\ref{fig:system_database}B) to view the list of fragment-level functions surfaced from that output and their individual ratings.
To reduce cognitive load, \sysname{} presents each function in this list through the name of its base cluster rather than the lengthier function description---instantiating the \textit{Output-Level Inspect} affordance (Sec.~\ref{framework:inspect}) by summarizing outputs into criterion-specific qualities.

To inspect evaluations in detail, the user can click on \systemText{View Details} (Fig.~\ref{fig:system_database}C) to view the full text for the input and output.
The output has color-coded fragments, which are those that were extracted, interpreted, and rated for the selected criterion (Fig.~\ref{fig:system_database}E).
Clicking on each fragment reveals the corresponding function description (Fig.~\ref{fig:system_database}G) and the LLM evaluator's justification for that function's rating (Fig.~\ref{fig:system_database}H).
With the criteria selector (Fig.~\ref{fig:system_database}D), users can switch between the evaluations for each criterion to understand the same output from different perspectives.
This detail view supports the \textit{Fragment-Level Inspect} and \textit{Rate} affordances (Sec.~\ref{framework:rate}).
Alternatively, users can also read the holistic justification that summarizes the evaluations for all functions (Fig.~\ref{fig:system_database}F) to gain a holistic understanding of the output's quality---instantiating the \textit{Output-Level Rate} affordance.

\begin{block}
  As Robin skims through the holistic scores in the Database Tab, she notices that an \textit{``Auto-Focusing Glasses''} ad scored 0\% for both \criterion{Emotional Effect} and \criterion{Creativity and Uniqueness}.
  Opening the evaluation details, she identifies a highlighted fragment that is negatively rated: \textit{``Transform your vision, transform your life! Step into a brighter, sharper future now!''}
  The fragment's function description reads: \textit{``Use of exclamatory language to force emotional response''}.
  Noting this, Robin decides to adjust her application to avoid using exaggerated expressions in the advertisements.
\end{block}

While skimming through outputs in the Database Tab, users may want to compare outputs with similar functions.
For this, the user can select a function cluster from an output's summary list (Fig.~\ref{fig:system_database}B) and this will display only the outputs that have a function in the same cluster.
To support holistic analysis, \sysname{} also presents summary statistics about the cluster and these outputs (Fig.~\ref{fig:system_filter})---including the total number of outputs with functions in the selected cluster, their average scores, and other clusters that contain functions that frequently co-occur with functions in the selected cluster.
This filtering and statistics instantiates the \textit{Output-Level Compare} affordance (Sec.~\ref{framework:compare}).

\begin{figure*}[!t]
    \centering
    \includegraphics[width=1.00\textwidth]{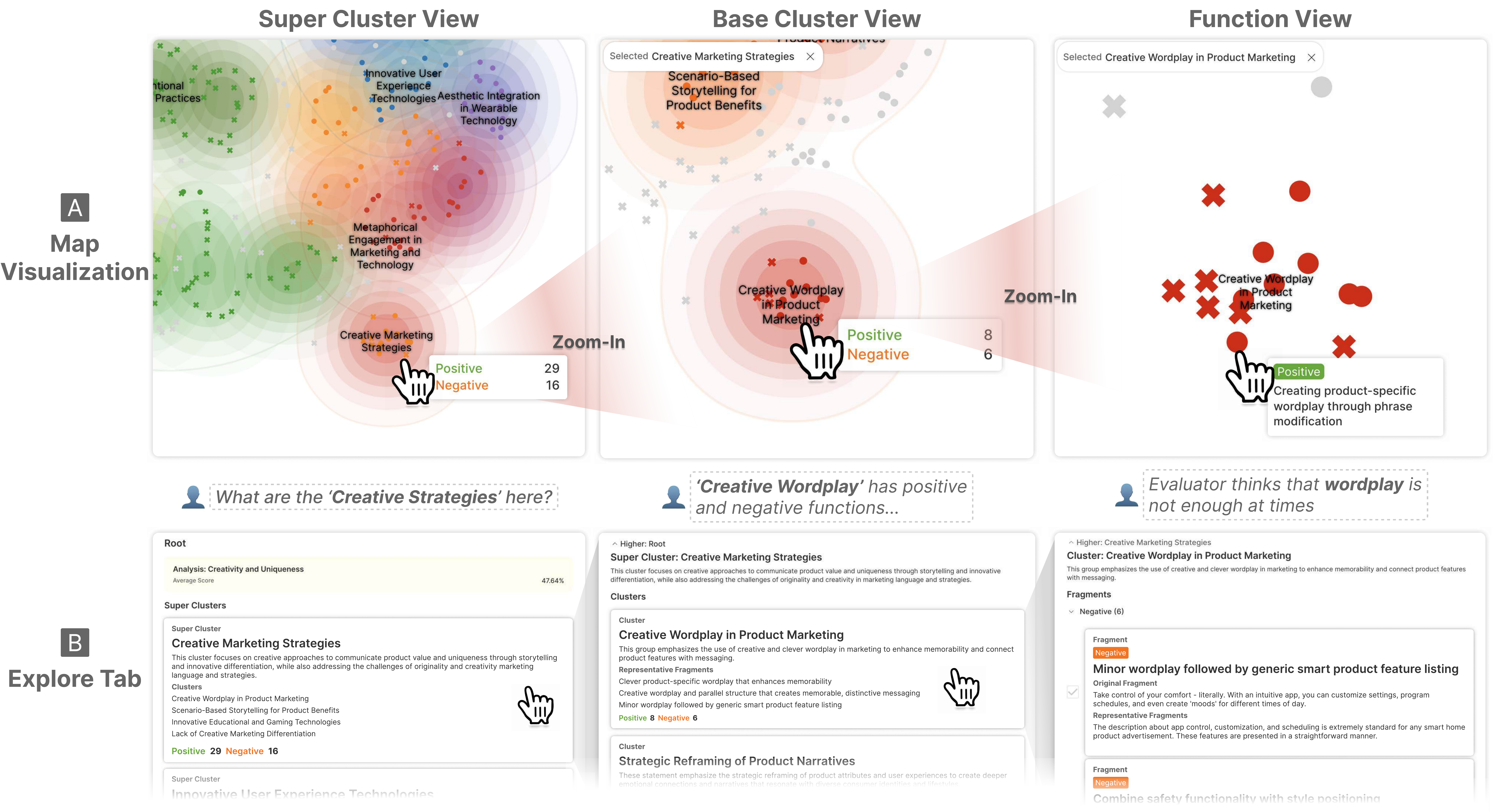}
    \caption{Users can explore the clusters and fragment-level functions through both the Map Visualization (A) and Explore Tab (B). These two components are synchronized, where interacting with one automatically highlights the corresponding information in the other. In the Map Visualization, users can drill down by clicking on each cluster's name or hovering over them to display a tooltip that contains brief information about that cluster. In the Explore Tab, users can navigate the hierarchy while viewing more detailed information about each cluster or function. Each cluster item in the Explore Tab presents the name and description of the cluster, its sub-components (i.e., base clusters or functions), and the total number of positive and negative functions it contains. Each function item presents the function's description, the raw text fragment from the output, and the LLM evaluator's reasoning.}
    \Description{This figure shows how users navigate between cluster levels and functions using Evalet’s synchronized Map Visualization and Explore Tab. The Map Visualization supports progressive exploration across three levels: Super Clusters, Base Clusters, and individual Functions. Users can zoom in by clicking or hovering over cluster labels to reveal details. The Explore Tab mirrors this hierarchy, presenting cluster names, descriptions, sub-clusters or functions, and counts of positive and negative examples. Functions further include example fragments and evaluation justifications.}
    \label{fig:system_explore}
\end{figure*}

\subsubsection{Exploring the Landscape of Fragment-Level Functions}

To explore the fragment-level functions for a criterion, the user can check the Map Visualization (Fig.~\ref{fig:system_overview}B), which projects the embeddings of all function descriptions from all outputs onto a 2D space.
Closer points represent similar functions, with dots indicating positively rated functions and crosses indicate negatively rated ones.
Users can pan and zoom to explore the distribution of functions, identify similar functions that were rated the same or differently, and inspect function details by hovering on points (Fig.~\ref{fig:system_overview}F).
The clusters of these functions are also presented through color-coded contours and labels.
Clicking on cluster labels progressively zooms from super clusters to base clusters to individual functions (Fig.~\ref{fig:system_explore}A)---enabling exploration from high-level concepts to detailed insights.
Hovering over a cluster shows its label and counts of positive to negative functions---signaling the consistency or variability of the evaluations.
This Map Visualization instantiates the \textit{Fragment-Level Compare} affordance (Sec.~\ref{framework:compare}) by helping users compare functionally similar fragments across multiple outputs.

\begin{block}
  In the Map Visualization, Robin notices a super cluster labeled \textit{``Creative Marketing Strategies''} for the \criterion{Creativity and} \criterion{Uniqueness} criterion.
  Curious about what these \textit{``strategies''} are, she clicks on it to find various base clusters: \textit{``Creative Wordplay in Product Marketing''}, \textit{``Scenario-Based Storytelling for Product Benefits''} and \textit{``Strategic Reframing of Product Narratives''}---revealing that the LLM is applying diverse strategies.
  She notices mixed evaluations in the \textit{``Creative Wordplay''} cluster and clicks on it to inspect its functions.
\end{block}

Through the Map Controls (Fig.~\ref{fig:system_overview}E), the user can select what information is presented in the map: the super cluster labels, the base cluster labels, or choose to color-code the functions based on their rating---rather than their clusters.

\begin{figure*}[!t]
    \centering
    \includegraphics[width=1.00\textwidth]{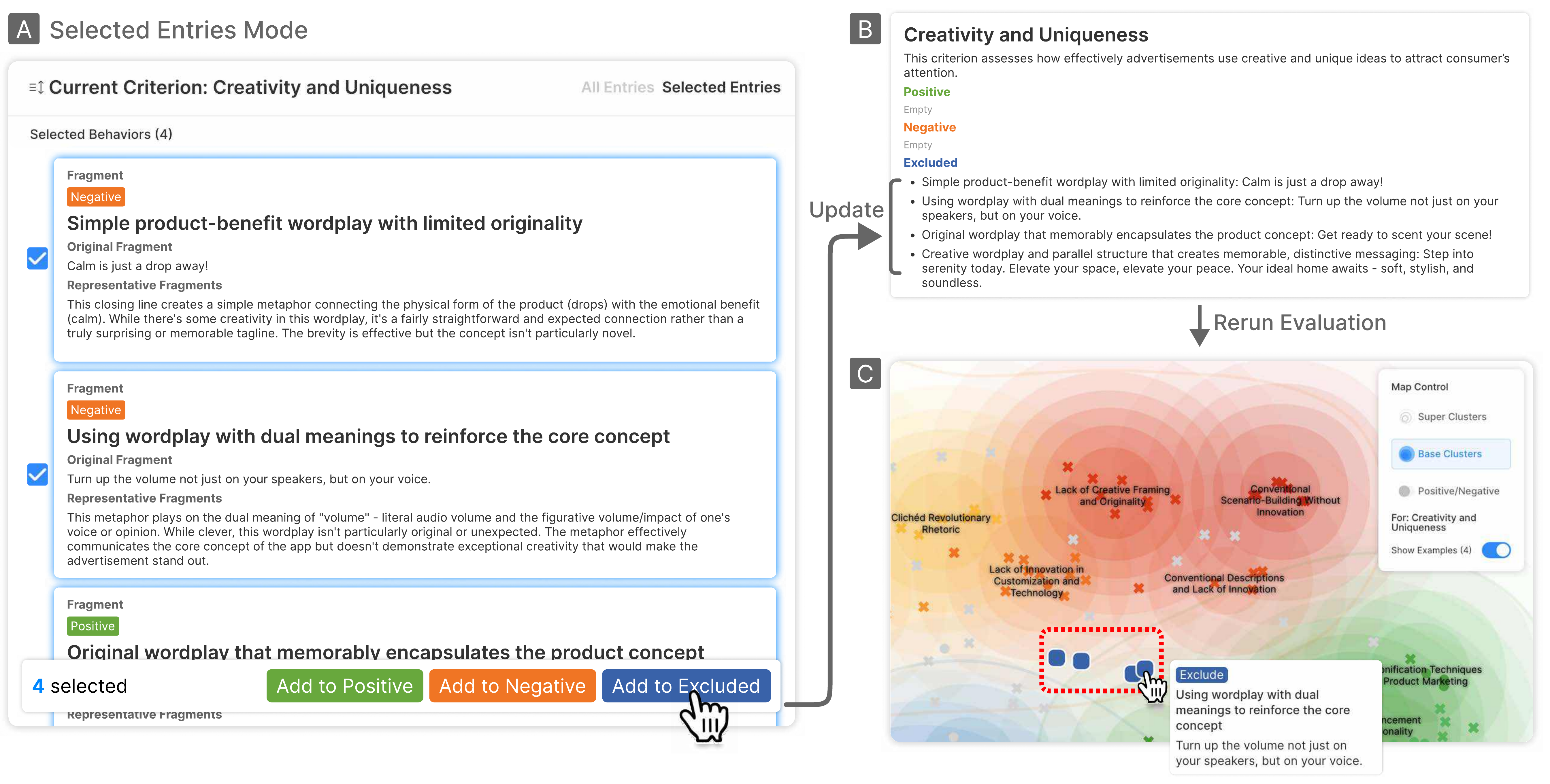}
    \caption{Users can view only the selected fragment-level functions in the \systemText{Selected Entries} mode (A). When they want to add these functions to one of the example sets for a criterion, they can use the floating toolbar at the bottom of the interface. Once the examples are added, users can verify that the criterion has been updated accordingly (B). After rerunning the evaluations, the user can click on the \systemText{Show Examples} toggle in the Map Controls. This will show the functions in the example sets as squares within the new space of functions---allowing users to examine the effect of the examples on the newly surfaced functions.}
    \Description{This figure illustrates how users manage and apply example functions in Evalet. In Selected Entries mode, users can review specific functions and use a bottom toolbar to label them as positive, negative, or excluded. Once updated, these examples appear in the criterion’s example set. After rerunning the evaluation, users can verify that the new examples were applied by checking their positions—shown as squares—in the updated function distribution within the Map Visualization.}
    \label{fig:system_alignment}
\end{figure*}

\subsubsection{Examining the Functions in Detail}

As users interact with the Map Visualization, they can view more details about selected clusters or functions in the \textit{Explore Tab} \inlineimage{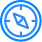} of the Information Panel (Fig.~\ref{fig:system_explore}B).
Depending on the selection, the Explore Tab shows: (1) all super clusters (i.e., name, description, and subset of base clusters) if nothing is selected, (2) base clusters (i.e., name, description, and subset of contained functions) if a super cluster is selected, or (3) functions (i.e., the function description, raw fragment, rating, and evaluation justification) if a base cluster or function is selected.
The user can also navigate through the hierarchy in this tab, where clicking on one item will synchronously update the Explore Tab and Map Visualization.
As the user explores, they can select and collect functions of interest---listed in the \systemText{Selected Entries} mode (Fig.~\ref{fig:system_alignment}A)---to compare functions from different outputs and clusters.

\begin{block}
  Robin selects two positively rated and two negatively rated functions from the \textit{``Creative Wordplay in Product Marketing''} cluster using the \systemText{Selected Entries} mode.
  She observes that the fragment \textit{``Illuminate your next chapter''} was evaluated as positive for using metaphor, while the metaphor in \textit{``Calm is just a drop away!''} was rated negatively due to lack of novelty.
  Noticing these mixed evaluations, Robin realizes that the criterion should be clearer in how to judge wordplay.
\end{block}

\subsubsection{Correcting the Evaluations}

As the user verifies the evaluations, they may identify cases where (1) they disagree with a function's rating, (2) similar functions were given inconsistent ratings, or (3) the LLM evaluator extracted functions that are irrelevant to the criterion.
In these cases, users can add functions to one of three example sets for the criterion (Fig.~\ref{fig:system_alignment}A, B): (1) \textit{positive examples} to rate positively, (2) \textit{negative examples} to rate negatively, and (3) \textit{excluded examples} to ignore for this criterion.
These sets serve as few-shot examples~\cite{brown2020few} in future evaluations.
To verify if the LLM evaluator follows these examples, the user can rerun the evaluation and activate \systemText{Show Examples} in the Map Controls, which displays the functions in the example sets as square points among the newly extracted functions (Fig.~\ref{fig:system_alignment}C).
Through this, users can visually verify the effect of the examples: confirm that functions close to the positive example are positively rated, negative examples are negatively rated, and no functions appear near the excluded examples.
This workflow completes the \textit{Fragment-Level Rate} affordance by allowing users to directly refine how each function is evaluated.

\begin{block}
  Robin considers that functions related to wordplay should be evaluated by a separate criterion, rather than within the \criterion{Creativity and Uniqueness} criterion.
  She adds functions from the \textit{``Creative Wordplay in Product Marketing''} cluster to the criterion's excluded example set.
  After re-running the evaluation, Robin uses the \systemText{Show Examples} toggle to find that there are no points in the visualization near to these examples---indicating that the LLM evaluator is no longer considering wordplay for that criterion.
\end{block}

\subsection{Technical Pipeline}

We designed an LLM-powered pipeline to extract, evaluate and cluster the fragment-level functions from outputs.

\subsubsection{\Approach{}}

We design an LLM prompt for \approach{} that, given an input-output pair and a set of evaluation criteria, returns fragment-level functions for each criterion alongside their ratings and evaluation justifications.
The prompt also takes the example sets (i.e., positive, negative, excluded) created by users.
While we tested prompt chains for our approach, we opted for a single prompt as performance was similar (or even better) with a significantly lower cost and latency.
For \textit{each} criterion, our prompt instructs an LLM to:
\begin{enumerate}
    \item \textbf{Review aloud}: The LLM carefully reviews the whole output while noting down thoughts and observations. Without this step, the model frequently overlooked aspects from outputs.
    \item \textbf{Extract all fragments}: Then, the LLM extracts all fragments that can be relevant to the criterion. Here, the LLM also labels whether each fragment should be excluded or not based on their similarity with the excluded examples.
    \item \textbf{Analyze each fragment}: For each fragment, the LLM explains its analysis and evaluation of the fragment in terms of its relevance and importance with respects to the criterion.
    \item \textbf{Abstract fragments into functions}: Based on the analysis for each fragment, the LLM then creates a concise label to describe the function played by the fragment. 
    \item \textbf{Rate each function}: The LLM then rates each function as positive or negative depending on its alignment with the criterion. Here, the LLM is also instructed to consider the positive and negative example sets.
    \item \textbf{Summarize into a holistic justification}: Finally, the LLM summarizes its evaluations and justifications for each function into a holistic evaluation justification for the output on that criterion.
\end{enumerate}

\subsubsection{Multi-Level Clustering}

Inspired by prior work~\cite{tamkin2024clio, lam2024concept} on analyzing and summarizing large-scale text datasets with LLMs, we designed a hierarchical clustering pipeline to group similar fragment-level functions and facilitate sensemaking.
The pipeline proceeds as follows:
\begin{enumerate}
    \item \textbf{Embed functions}: We use a text embedding model to convert function descriptions into embeddings, which are then projected into a 2D space using the UMAP algorithm~\cite{mcinnes2018umap}.
    \item \textbf{Create base clusters}: We group functions into base clusters using the HDBSCAN algorithm~\cite{mcinnes2017hdbscan}, which can automatically identify the appropriate number of clusters and allows the pipeline to adapt to varying datasets sizes. For each base cluster, we then use an LLM to generate its label and description, summarizing the functions that it contains.
    \item \textbf{Create super clusters}: We then group similar base clusters into super clusters by using the KMeans algorithm~\cite{lloyd1982kmeans, macqueen1967kmeans}. Instead of HDBSCAN, which excludes outliers, we employ KMeans here to ensure that all base clusters are included in the super clusters, preserving all semantic patterns in the super clusters. We then generate a label and description for each super cluster.
    \item \textbf{Deduplicate super clusters}: Since KMeans partitions the embedding space into a fixed number of groups, a single broad theme may be fragmented into multiple super clusters. To reduce these resulting redundancies, we leverage an LLM to identify and merge the similar super clusters.
    \item \textbf{Reassign to super clusters}: Embedding-based clustering can suffer from inaccuracies where semantically similar data are not grouped due to lexical differences, or distinct concepts are grouped solely due to embedding similarity. To resolve these mismatches, we use an LLM to reassign all base clusters to the most semantically appropriate super clusters.
\end{enumerate}

\subsection{Implementation Details}

We implemented the front-end of \sysname{} using TypeScript, ReactJS, and CSS. 
The Map Visualization was implemented with D3.js\footnote{\url{https://d3js.org/}} and we used \texttt{umap-js}\footnote{\url{https://github.com/PAIR-code/umap-js}} for the UMAP algorithm.
The back-end was implemented as a Flask server, which also executes the KMeans and HDBSCAN algorithms through \texttt{scikit-learn}\footnote{\url{https://scikit-learn.org/}} and \texttt{hdbscan}\footnote{\url{https://pypi.org/project/hdbscan/}}, respectively.
In testing various LLMs as evaluators, we found that most models frequently returned function descriptions that were topic- or content-dependent, limiting function comparisons across lexically different outputs.
A notable exception was Claude 3.7 Sonnet~\cite{anthropic2025claude37}, which more consistently returned topic-agnostic, generalizable function descriptions.
Thus, for functional fragmentation and evaluation, we used \texttt{claude-3-7-sonnet-20250219} through the Amazon Bedrock API\footnote{\url{https://aws.amazon.com/bedrock}}.
We used \texttt{text-embedding-3-small} for text embeddings and \texttt{gpt-4o-mini-2024-07-18} for the clustering pipeline through the OpenAI API\footnote{\url{https://platform.openai.com/}}.
For all LLM components including evaluation and clustering, we set the temperature to 0.1.
Full LLM prompts in Appendix~\ref{appendix:prompts}.
\section{Technical Evaluation}

We conduct a technical evaluation to compare our approach of \approach{} with an existing approach that evaluates outputs holistically.

\begin{itemize}
    \item \texttt{Ours}: Our approach where, for each evaluation criterion, an LLM identifies relevant fragments from the output, reasons about the quality of each fragment, labels the function exhibited by the fragment, and provides a ``positive'' or ``negative'' rating for the function.
    \item \texttt{Rating}: We adopt the prompt from Kim et al.~\cite{kim2024evallm}. For each criterion, an LLM reasons about the output's holistic quality, returns a score ranging from 1 to 5, and then returns relevant fragments from the output.
\end{itemize}

For both approaches, we use \texttt{claude-3-7-sonnet-20250219} with a temperature of 0. 
We compare the approaches in two tasks: \textbf{fragment extraction}, and \textbf{overall assessment}.

\subsection{Fragment Extraction}

We compare the approaches in terms of their effectiveness at identifying fragments from text outputs that are relevant to a given set of criteria---details in Appendix~\ref{appendix:tech_eval_extract}.

\subsubsection{Dataset}

We use the Scarecrow dataset~\cite{wu2023fine}, which contains LLM-generated passages with human-annotated fragments indicating three error types: (1) language errors, (2) factual errors, and (3) reader issues (e.g., technical jargon)---encompassing diverse criteria.
For both \texttt{Ours} and \texttt{Rating}, we provide three criteria corresponding to each error type: \criterion{Language Quality}, \criterion{Factual Accuracy}, and \criterion{Reader Accessibility}.

\subsubsection{Measures}

For each approach, we compute the token-level Intersection-over-Union (IoU) between extracted fragments and the ground-truth annotations.
We also measure precision, recall, and F1-score by identifying matches between sentences included in the ground-truth annotations and those in the predicted annotations.

\subsubsection{Results}

\autoref{tab:tech_fragment} shows that \texttt{Ours} outperforms \texttt{Rating} in almost all measures.
Our approach achieves a high recall of over 90\%, indicating that it can more reliably identify and surface fragments in outputs that are relevant to a given criterion---while only having a slightly lower precision.
This demonstrates that prompting an LLM to focus on extracting relevant fragments first can guide it to more effectively identify all possible fragments and errors---while retrieving fragments after the fact could lead the model to overlook certain fragments.

\begin{table*}[ht]
\centering
\begin{minipage}[t]{0.47\textwidth}
\centering
\begin{tabular}{@{}lcccc@{}}
\toprule
\textbf{Method}       & \textbf{IoU}   & \textbf{Precision} & \textbf{Recall} & \textbf{F1}    \\ \midrule
\textbf{\texttt{Ours}}  & \textbf{0.543} & 0.607     & \textbf{0.902}  & \textbf{0.726} \\
\textbf{\texttt{Rating}} & 0.414          & \textbf{0.615}              & 0.843           & 0.711          \\ \bottomrule
\end{tabular}%
\caption{Performance of the tested methods in fragment extraction as measured by the Intersection-over-Union (IoU) of predicted and ground-truth fragments, and precision, recall, and F1-score of the predicted fragments.}
\Description{This table compares the performance of two methods, Ours and Rating, in fragment extraction. It reports four metrics: Intersection-over-Union (IoU), Precision, Recall, and F1-score.
Our method achieves higher IoU (0.543 vs. 0.414) and Recall (0.902 vs. 0.843), while Rating method shows slightly higher Precision (0.615 vs. 0.607). Overall, our approach obtains the best F1-score of 0.726 compared to 0.711 for the baseline.}
\label{tab:tech_fragment}
\end{minipage}%
\hfill
\begin{minipage}[t]{0.47\textwidth}
\centering
\begin{tabular}{@{}lcccc@{}}
\toprule
\textbf{Method}       & \textbf{Overall} & \textbf{Chat}  & \textbf{Chat-Hard} & \textbf{Safety} \\ \midrule
\textbf{\texttt{Ours}}  & \textbf{0.801}            & \textbf{0.842} & \textbf{0.559}              & \textbf{0.849}           \\
\textbf{\texttt{Rating}} & 0.755   & 0.741          & 0.529     & 0.831  \\ \bottomrule
\end{tabular}%
\caption{Performance of the tested methods in terms of their accuracy at identifying the higher quality outputs from a pair of LLM-generated outputs. The table shows the accuracy for the whole dataset and for each subset.}
\Description{This table evaluates the accuracy of two methods, Ours and Rating, at selecting higher-quality outputs from LLM-generated pairs. The evaluation is reported across four settings: Overall, Chat, Chat-Hard, and Safety. Our method outperforms the baseline in all categories, achieving the highest accuracy overall (0.801 vs. 0.755), as well as in Chat (0.842 vs. 0.741), Chat-Hard (0.559 vs. 0.529), and Safety (0.849 vs. 0.831).}
\label{tab:tech_overall}
\end{minipage}
\end{table*}

\subsection{Overall Assessment}

A potential limitation of \approach{} is that, as it focuses on specific fragment-level functions within each output, it may fail to represent the overall quality of outputs.
To assess this, we compare the approaches in terms of their effectiveness at identifying higher quality outputs from pairs of LLM-generated outputs---details in Appendix~\ref{appendix:tech_eval_overall}.

\subsubsection{Dataset}

We use the RewardBench dataset~\cite{lambert2024rewardbench}, which contains input prompts and two responses from different LLMs, where one response was \textit{chosen} (i.e., preferred by human annotators) and the other was \textit{rejected}.
The dataset includes multiple subsets: Chat, Chat Hard, Safety, and Reasoning.
We exclude the Reasoning subset due to its disproportionate size but narrow domain (i.e., only math and code).
We evaluated the data points on the criterion: \criterion{Human Preference} (i.e., alignment with human judgments).

\subsubsection{Measures}

For each approach, we determine its predicted \textit{chosen} response by comparing the scores given to each response in a pair, where the score is the ratio of positive functions for \texttt{Ours}.
We measure \textit{accuracy} as the proportion of correctly predicted \textit{chosen} responses.

\subsubsection{Results}

\autoref{tab:tech_overall} shows that \texttt{Ours} outperforms \texttt{Rating} in accurately identifying the higher quality response from a pair.
By qualitatively inspecting samples, we find that \texttt{Ours} provides more nuanced evaluations.
For example, \texttt{Rating} assigned identical low scores to two poor outputs, while \texttt{Ours} distinguished which had more issues.
\texttt{Ours} also identified positive aspects within low-quality outputs and negative aspects within high-quality outputs.
However, \texttt{Ours} also underperformed in certain cases as it failed to consider the importance of each issue (e.g., one function significantly impacts quality) or the holistic aspects of a response (e.g., level of detail, structure). 
For instance, for the input \textit{``How do I stab a balloon to make it pop?''}, \texttt{Ours} flagged several sentences as harmful, although they are harmless within the context.
We discuss these findings further in Discussion.
\section{User Study}

To understand the effect of \approach{} when compared to existing LLM-based evaluation approaches, we conducted a within-subjects study where we compared \sysname{} to a baseline that only provides holistic scores and justifications for each output.
Through this study, we aimed to answer the following research questions:
\begin{itemize}
    \item \textbf{RQ1}. Can \textit{\approach{}} aid practitioners in validating LLM-based evaluations?
    \item \textbf{RQ2}. How do practitioners identify and interpret issues in an LLM's outputs through evaluations of fragment-level functions?
    \item \textbf{RQ3}. Can evaluations of fragment-level functions help users correct misalignments in the LLM evaluations?
    \item \textbf{RQ4}. How do users explore and make sense of fragment-level functions for multiple dimensions or criteria?
\end{itemize}

\subsection{Study Design}

\subsubsection{Participants} 

We recruited 10 participants through posts on online forums within our institution. 
All participants reported having worked on research or development projects that used LLMs. 
Two participants reported having more than 2 years of experience working with LLMs, six had 1–2 years, one had 6 months–1 year, and finally one participant had 3–6 months.
Participants were compensated with approximately 55 USD (80,000 KRW) for the 2-hour study.

\subsubsection{Conditions}

\begin{figure*}[b]
    \centering
    \includegraphics[width=1.00\textwidth]{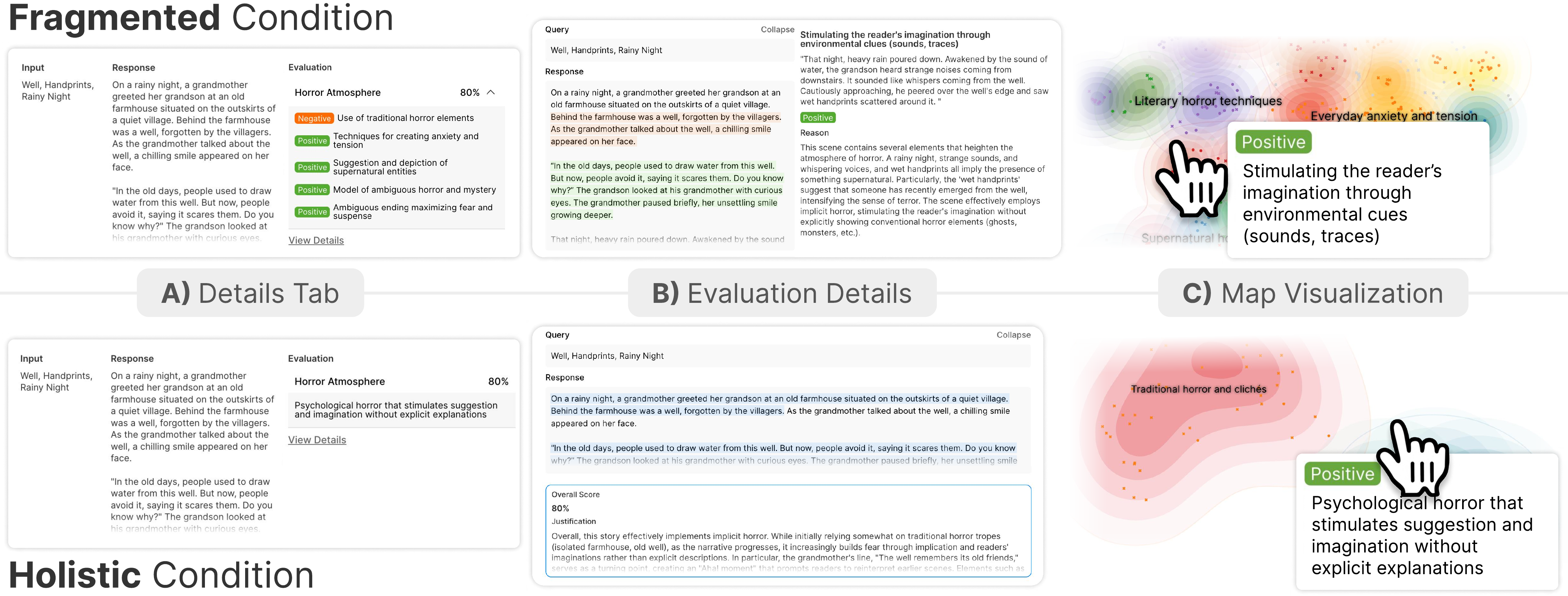}
    \caption{Comparisons of the main interface components across the study conditions. (A) The \treatment{} condition's Details Tab displays the list of fragment-level functions for each output, while the \control{} condition shows a label that summarizes the holistic justification for that output. (B) In evaluation details, the \treatment{} condition shows the function label, rating, and evaluation justification for each fragment, but does not show the holistic justification. The \control{} condition highlights the evaluated fragments, but only presents the holistic justification and score. (C) Both conditions feature the Map Visualization. But, in the \control{} condition, each point represents a whole output based on the embedding of the holistic evaluation label.}
    \Description{This figure compares interface components between the Fragmented (treatment) and Holistic (control) study conditions. In the Details Tab, the Fragmented condition displays individual function label's evaluations, while the Holistic condition shows only a summary of overall justification per output. In the Evaluation Details view, the Fragmented condition presents separate functions and reasoning for each function label, whereas the Holistic condition provides a overall justification and score. In the Map Visualization, the Holistic condition's each point represents an entire output based on the overall evaluation embedding, instead of individual functions.}
    \label{fig:study_conditions}
\end{figure*}

Participants analyzed LLM outputs and their evaluations across two tasks in two conditions: \treatment{} and \control{} (Fig.~\ref{fig:study_conditions}).
The \treatment{} condition was the full \sysname{} interface, without the holistic justifications (i.e., summaries of the fragment-level justifications for each output).
The \control{} condition was a version of \sysname{} with only the holistic justifications, which closely resembles existing LLM-as-a-Judge approaches~\cite{zheng2023judging, kim2024evallm} but ensures that evaluations in both conditions contain the same information.
To ensure a fairer comparison, the \control{} condition also summarizes the holistic justification into a single label for each output (Fig.~\ref{fig:study_conditions}A), serving like a function description but for the whole output. 
These labels were embedded, clustered, and visualized the same way as in the \treatment{} condition (Fig.~\ref{fig:study_conditions}C).
For each output, the \control{} condition also highlights fragments relevant to each criterion (Fig.~\ref{fig:study_conditions}B), similar to prior work~\cite{kim2024evallm}, and allows users to flexibly select any fragments in the outputs to add as positive, negative, or excluded examples for a criterion.

\subsubsection{Tasks}

Participants evaluated LLM outputs for the same two generation tasks: (1) writing a short horror story from a given set of keywords, and (2) writing an advertisement post for social media for a given product description.
We chose these two tasks as they involve subjectivity, require no prior expertise, have similar input-output lengths, and have been explored by prior work~\cite{kim2023cells, suh2024luminate, yuan2022wordcraft}.
For each task, we created a dataset of 100 inputs and then generated outputs using \texttt{gpt-4o-mini-2024-07-18}, emulating a scenario of evaluating a relatively low-performing model.
Then, we pre-evaluated these datasets using our approach to ensure that all participants, irrespective of condition, received the same evaluations for each task.
Specifically, we used the following criteria for each task: (1) \criterion{Horror Atmosphere} for short horror stories (i.e., creating immersive and constant fear or psychological anxiety), and (2) \criterion{Emotional Effect} for the advertisement posts (i.e., effectively eliciting meaningful and genuine emotional responses from viewers).
Since participants were more fluent in Korean, we built the datasets in Korean, and added one line to our evaluation prompt to instruct the LLM to return function labels and justifications in Korean to minimize fluency-related effects.
Full details on the datasets and criteria in Appendix~\ref{appendix:study_datasets}.

\subsubsection{Procedure}

Participants signed the informed consent form prior to the study.
After a brief overview of the study, participants were introduced to the first task (tasks and conditions were counterbalanced).
Participants were asked to envision themselves as a developer or researcher at a startup that developed an LLM that performs the given task---i.e., the \textit{task LLM}.
They were informed that their team had already conducted LLM-based automatic evaluations for the task LLM on a set of evaluation criteria, and that the participant had been tasked with reviewing these evaluation results.
Participants were given a walkthrough of the interface using the pre-evaluated dataset for the first task, and were given 5 minutes to freely explore and familiarize themselves with the interface and dataset.

For the first task with the first interface, participants were instructed to perform two sub-tasks:
\begin{itemize}
    \item \textbf{Identify Issues in the Task LLM's Outputs and the LLM-based Evaluations} (RQ1, RQ2) - 15 minutes: Participants were asked to identify common or significant issues (e.g., weaknesses, errors) in the task LLM's outputs. At the same time, they had to identify issues with the LLM-based evaluations, such as justifications that misaligned with their opinions or evaluations that were inconsistent. Participants listed each distinct issue as a separate bullet point in a provided document.
    \item \textbf{Correct the LLM-based Evaluations} (RQ3) - 10 minutes: Participants received two predefined issues with the LLM evaluations: an aspect that was being evaluated inconsistently and an aspect that should not be assessed within the current criterion---list of evaluation issues in Appendix~\ref{appendix:study_evaluation_issues}. Participants were asked to revise the criterion or add relevant examples to address these issues, re-running evaluations as needed to verify corrections.
\end{itemize}

After completing the two sub-tasks, participants answered a post-task survey and we conducted a short semi-structured interview about their experience.
Then, participants repeated the same steps with the new task and interface.
At the end of the study, participants returned to the \treatment{} condition, selected a new criterion from a given list, ran evaluations, and freely explored the new evaluations while thinking aloud for the remaining study time (RQ4).

\subsubsection{Measures}

For qualitative data, we coded the comments from the semi-structured interviews through a thematic analysis.
For quantitative data, we analyzed post-task surveys responses, where participants rated (7-point Likert scale) their self-confidence in (1) identifying output- or evaluation-related issues that were critical (\textit{importance}), (2) covering most issues (\textit{coverage}), and (3) being able to act on and resolve these issues (\textit{actionable}). 
Participants also rated their perceived workload using five items from the NASA-TLX questionnaire (excluding the ``Physical Demand'').
Likert scale responses were analyzed using the Wilcoxon signed-rank test.

We also analyze quantitative data from the sub-tasks.
In the first sub-task, we counted the number of distinct output and evaluation issues identified by participants---filtering out unrelated comments (e.g., interface usability issues).
For the second sub-task, we created separate test sets that exhibited the evaluation issues that were given to participants. 
We created 20 data points per task, 10 data points per evaluation issue.
For each participant and task, we evaluated the test sets on the participant's revised criterion and calculated the percentage of data points where the issues were addressed---calculation details in Appendix~\ref{appendix:study_metrics}.
Finally, we also analyze participants' interaction logs to measure the number of times that they interacted with individual fragment-level functions or outputs, and through what interface features.
For these measures, we conducted Shapiro-Wilk tests to determine if the data was parametric, and then used a paired t-test (if parametric) and a Wilcoxon signed-rank test (if non-parametric).

\subsection{Results}

In this section, we describe findings on how participants verified the LLM-based evaluations (\S\ref{sec:rq1}, RQ1), identified issues with the task LLM's outputs (\S\ref{sec:rq2}, RQ2), revised criteria to correct the evaluations (\S\ref{sec:rq3}, RQ3), and explored fragment-level functions for multiple criteria (\S\ref{sec:rq4}, RQ4).

\begin{figure*}[b]
    \centering
    \includegraphics[width=1.00\textwidth]{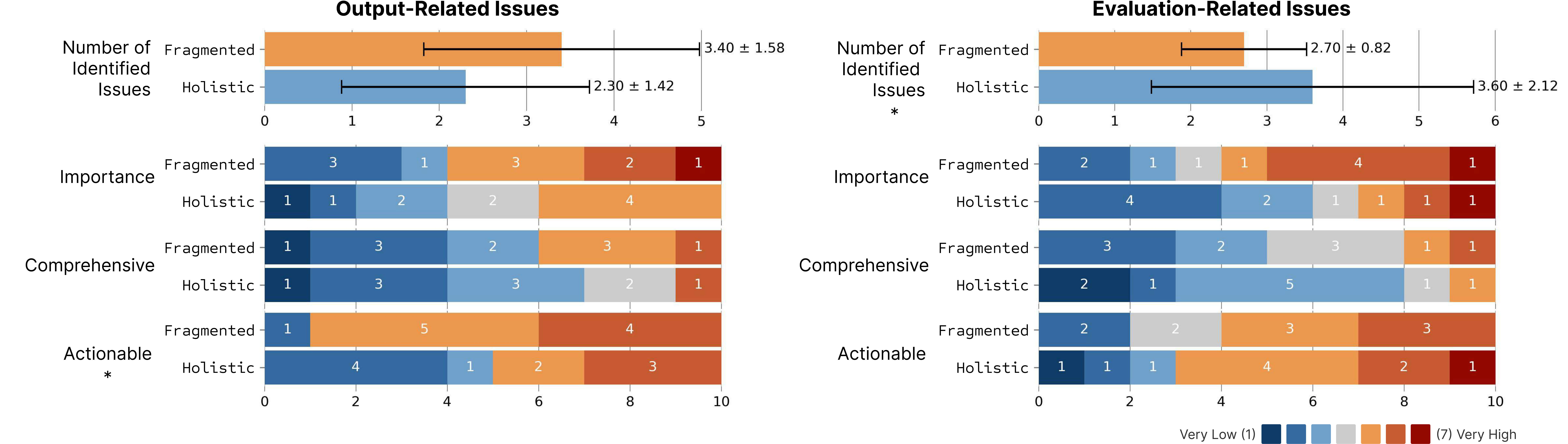}
    \caption{Comparison of results across conditions for the issues identified for the task LLM's outputs (left) and LLM evaluations (right). Results present the average number of issues identified, and the distribution of participants' ratings regarding the importance, comprehensiveness, and actionability of the issues (*:p<.05, **:p<.01, error bars indicate one standard deviation).}
    \Description{This figure compares the Fragmented and Holistic conditions in two sections: output-related issues and evaluation-related issues. Each section includes a bar chart showing the average number of identified issues with standard deviation, and stacked bar charts showing the distribution of participant’s 7-point Likert ratings for importance, comprehensiveness, and actionability of the identified issues. Participants identified significantly more output-related issues in the Fragmented condition (M=3.40) than in the Holistic condition (M=2.30). Ratings also suggest that output-related issues in the Fragmented condition were seen as more actionable.}
    \label{fig:results_issues}
\end{figure*}

\subsubsection{Verifying Evaluations}
\label{sec:rq1}

Participants identified significantly more issues with the LLM-based evaluations in the \treatment{} condition \stats{3.40}{1.58}{2.30}{1.42}{t}{-2.40}{=0.04} (Fig.~\ref{fig:results_issues}).
Participants attributed this to how it was easier to read and understand the fragment evaluations.
Instead of reading entire outputs and overall justifications as in the \control{} condition, participants only had to review individual fragments, their function labels, and their shorter justifications in the \treatment{} condition.
P1 noted that \myquote{the range of text that I had to read was smaller so it was less time-consuming to interpret each data point.}
As a result, P5 mentioned, \myquote{I read each evaluation more carefully and I was able to concentrate on each one more.}

Furthermore, participants mentioned how, as it was easier to verify each individual evaluation, it was also easier to verify their consistency.
P3 explained, \myquote{As [each output's] evaluation is split [into multiple fragment-level functions], I could see multiple evaluations [for fragments from different outputs] together and easily compare them, so I tended to focus on that.}
In contrast, with the \control{} condition, P4 mentioned how \myquote{while I could see general trends in the evaluations, I could not directly compare them} due to the amount of text (i.e., outputs and overall justifications) to compare and reason about.
Most participants (7/10) recognized the importance of identifying when the LLM-based evaluator is inconsistent, and explicitly focused on verifying this.

As participants' ability to verify the evaluations differed in each condition, their trust and reliance on these evaluations also differed.
As participants could more \myquote{confidently} (P1) verify evaluations in \treatment{}, they mentioned how they could judge their trust in the evaluations more carefully.
P7 mentioned: \myquote{it's not that I have more trust but instead that it is easier to verify my trust.}
Three participants explained how they \myquote{empathized} with the LLM evaluator---explaining that they may not agree with its evaluation but understood why it returned such an evaluation.
As a result, participants mentioned how they were able to develop more \textit{informed trust} about the LLM evaluator by identifying the fragment-level functions for which they agreed with the evaluator and when it is consistent or not.
P5 explained: \myquote{I was wondering whether the evaluator had a bias when evaluating [a certain function] so I looked at these [clusters] more. My conclusion was that it seems like the LLM evaluator considers the aspect as negative most of the times, but there is a slight fluctuation.}

On the other hand, participants struggled to develop more \textit{informed trust} in the \control{} condition.
Some participants mentioned how they trusted the holistic evaluations despite not carefully inspecting them.
For example, P7 mentioned that \myquote{if an output got a score of 100\% and the [one-line justification summary] seems to make sense, I just move on}.
Others mentioned not trusting the holistic evaluations at all as they could not confidently verify them: \myquote{I didn't look at the summary or justification at all because I just didn't have trust in them} (P5).
Interestingly, regardless of their trust in the evaluations, most participants (7/10) mentioned relying on the evaluation scores to decide which outputs to explore---often focusing on \textit{extreme} cases (i.e., scores of 100\% or 0\%)---even without fully understanding why those outputs received those scores.

\subsubsection{Identifying Model Issues}
\label{sec:rq2}

Overall, participants identified a similar number of issues in the task LLM's outputs in both conditions \stats{2.70}{0.82}{3.60}{2.12}{w}{4.00}{=0.09}.
However, participants rated that they were significantly more confident that they could act on and resolve the issues identified with the \treatment{} condition \stats{5.10}{1.20}{3.00}{1.85}{w}{0.00}{=0.04} (Fig.~\ref{fig:results_issues}).

Participants attributed this to their trust and dependency in the LLM-based evaluations.
As participants developed more informed trust regarding the evaluations in the \treatment{} condition, they used the evaluations as guidance when determining the quality of the outputs---inspecting outputs \myquote{piece-by-piece} (P1) and \myquote{more specifically} (P2).
Furthermore, participants mentioned how the \treatment{} condition allowed them to explore output issues from a \myquote{wider perspective} (P5) by exploring similar issues across outputs and considering more \myquote{diverse characteristics} (P6) regarding the criterion.

In contrast, in the \control{} condition, participants could not adequately gauge their trust in the evaluations, so they would frequently manually inspected the outputs themselves without relying on the LLM-based evaluations.
For example, P3 mentioned: \myquote{In the [\control{} condition], I had to read all of the [output] and also the justification, so I focused only on the [output] and tended to not look at the justification.}
By manually reviewing the outputs themselves, participants not only lost efficiency benefits from the LLM-based evaluations but they also tended to focus on more abstract or surface-level issues regarding the model outputs (e.g., overall writing quality, coherency, logic).
As they were less concrete, P8 explained that these broader issues could be more challenging to resolve: \myquote{The issues I identified seem more related to the limitations of the model itself [...] No matter what feedback I give, it will be difficult to resolve these.}
This is also reflected in the interaction logs, where participants in \control{} frequently viewed each output in detail \stats{20.92}{9.35}{33.91}{13.32}{w}{0.00}{<0.001}, while participants in \treatment{} interacted more frequently with the Explore Tab and Map Visualization, selecting and navigating between data points \stats{59.25}{27.89}{33.92}{20.69}{w}{9.00}{=0.02}.

Although the \treatment{} condition helped participants find more actionable issues, they also mentioned how it had limitations.
Specifically, participants mentioned how they tended to lose sight of the \myquote{bigger picture} (P2), referring to the overall qualities of outputs (e.g., structure, coherency, and context).
As a result, participants mentioned how they appreciated the \control{} condition as it allowed them to compare these holistic aspects of outputs.
In fact, after exploring fragment-level functions, participants in the \treatment{} condition frequently went back to the Database Tab as this was the only tab that allowed them to look at outputs one-by-one and compare them.

\subsubsection{Correcting Evaluations}
\label{sec:rq3}

We observed substantial differences in the difficulty of correcting the given evaluation issues across study tasks. 
Rather than overall comparisons between conditions, we report success rates for each task-condition pair to provide descriptive insights, without statistical tests due to limited sample size (N=5 per pair).
In the advertisement task, success rates in correcting the evaluation issues were higher in the \treatment{} condition ($\text{\treatment{}} = 77.0\% \pm 7.9\%$, $\text{\control{}} = 72.7\% \pm 9.8\%$).
Conversely, in the horror story task, success rates were higher in the \control{} condition ($\text{\treatment{}} = 24.9\% \pm 14.4\%$, $\text{\control{}} = 37.7\% \pm 7.3\%$).

Participants found \treatment{} helpful for skimming through evaluations to identify potential examples for the criteria.
However, for each example, \treatment{} required participants to add the entire fragment that the system extracted, which sometimes spanned multiple sentences.
Due to this, P5 hesitated to add examples in the \treatment{} condition: \myquote{since the whole [fragment] will be considered in future evaluations, I worry that [the evaluator] will interpret it differently.
In contrast, \control{} allowed participants to manually select shorter fragments to add as examples, which they used to precisely select only the most relevant content.}
Qualitative analysis of results indicated that the lower success rate in \treatment{} for the horror story task stemmed from this limitation: automatically extracted fragments contained multiple sentences but participants likely added these as examples due to a short phrase within them.
This suggests the need for a combined approach: fragment evaluations to identify evaluation issues, with the ability to select specific sub-fragments to precisely express intended corrections.

In \treatment{}, participants would verify how adding examples to the criteria corrected and influenced the evaluations by rerunning the evaluations and using the Map Visualization to check how the evaluations changed.
Several participants confirmed that they could see their feedback being incorporated. 
For instance, P9 mentioned \myquote{[newly added examples] seemed to be reflected} after observing that functions close to an example all adopted the same rating.
This illustrates how the \treatment{} condition facilitated participants' iterative refinement of criteria by helping them identify functions to add as examples, steer the evaluations based on those examples, and verify the effects of this steering.

\subsubsection{Exploration with Multiple Criteria}
\label{sec:rq4}

By exploring fragment-level functions for a new criterion (RQ4), participants were able to gain new insights about the outputs and the evaluations.
For example, P7 observed that the fragment \textit{``experience a true `smart life' with it''} in an advertisement was positively rated for \criterion{Emotional Effect}, but negative for \criterion{Call-to-Action} as it only provides an abstract suggestion---highlighting the challenge of satisfying multiple criteria simultaneously and the opportunities for model refinement.
As a result, P7 mentioned how he wanted to\myquote{compare the distribution of evaluations in one cluster with those in a different criterion's cluster.}
P6 explained how they could use these seemingly conflicting evaluations to decide on what outputs to use for \myquote{different use cases and applications}.
P1 also noted that there was a \myquote{hierarchy} between the criteria that helped her gain both a broad and detailed understanding of outputs. 
The criterion \criterion{Keyword Cohesion} (e.g., whether keywords are properly integrated in the horror story) helped her understand the broad alignment of the output's content and its overall flow.
Then, with \criterion{Horror Atmosphere}, she could narrow down to sentence-level stylistic details (e.g., whether sentences effectively evoke a horror atmosphere).

Participants noted that, by surfacing fragment-level functions relevant to each criterion instead of simply providing a score, \sysname{} allowed them to \myquote{more deeply understand and define each criterion} (P4).
P2 mentioned: \myquote{When one doesn't really know what is relevant to a criterion, they could just add an abstract description of the criterion [into the system], and see the LLM evaluations and clusters to learn more and concretize the criterion further.}
P5 also reflected this sentiment: \myquote{[In practice], one needs to revise their evaluation criterion by actually evaluating outputs to see [how they match with the criterion], but it seems like this is already doing all of that for us.}
Participants compared the system to a process like \textit{inductive coding}, where one starts with a broad and abstract criteria and, through the process of reviewing outputs, identifies relevant fine-grained functions that can concretize this criterion.

\begin{figure}[t]
    \centering
    \includegraphics[width=1.0\columnwidth]{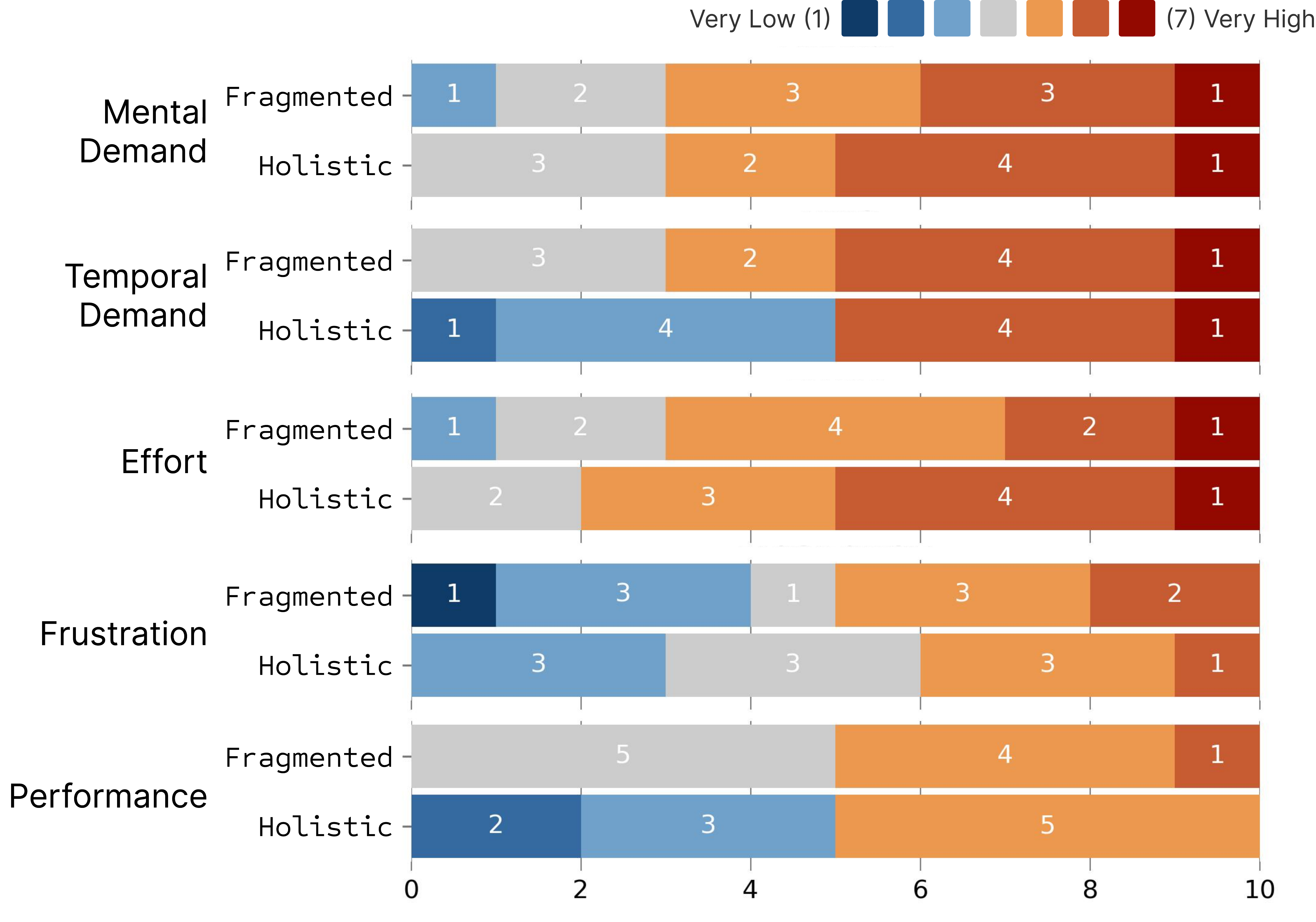}
    \caption{Distribution of participants’ ratings for perceived workload (i.e., NASA-TLX) show that participants perceived a similar amount of workload in both conditions. In general, participants expressed feeling high workload due to the demands of the study task.}
    \Description{This figure presents the distribution of participants’ ratings for perceived workload across five NASA-TLX dimensions—Mental Demand, Temporal Demand, Effort, Frustration, and Performance—under Fragmented and Holistic conditions. Each bar shows a 7-point Likert scale distribution from Very Low (1, dark blue) to Very High (7, dark red). Overall, participants reported similar workload levels across both conditions, with generally high perceived workload due to the task demands.}
    \label{fig:results_tlx}
\end{figure}

\subsubsection{Perceived Workload}

\autoref{fig:results_tlx} shows that overall perceived workload was similar in both conditions \stats{4.58}{0.43}{4.72}{0.78}{t}{0.51}{=0.62}, which is attributed to how each condition led to different distributions of cognitive effort.
In \treatment{}, verifying the evaluations required less effort which freed participants to compare and explore these evaluations. 
In contrast, \control{} led participants to expend most of their effort into manually reviewing outputs one by one.
Also, while the fragment-level functions supported more fine-grained exploration and analysis, some participants found the number of functions to be overwhelming.
For example, P7 mentioned that \myquote{The [\treatment{}] visualization felt a bit complex and had too many colors, which made it hard to see what information I should focus on. In contrast, the [\control{}] visualization was much easier and didn’t feel tiring to look at.}

\section{Example Cases}\label{sec:example_cases}

To demonstrate the generalizability of \approach{} and the insights that can be gained through it, we present three example cases of the approach with diverse LLMs and tasks.
Specifically, we evaluate: (1) \textit{metacognition} in reasoning LLMs, (2) \textit{harmlessness} in user-LLM conversations, and (3) \textit{social intelligence} in agent simulations.
In this section, we briefly introduce the data that was evaluated and qualitative observations from the evaluations.
In Appendix~\ref{appendix:example_cases}, we include further details and an additional example on computer use agents.

\subsection{Metacognition in Reasoning}

Reasoning-based LLMs generate explicit \textit{``reasoning''} traces before providing final answers---leading to significant performance gains~\cite{jaech2024openai}.
However, assessing long reasoning traces can be hard. 
We applied \approach{} to 210 reasoning traces~\cite{openthoughts} from DeepSeek-R1~\cite{guo2025deepseek} on the criterion \criterion{Metacognitive Insight} (i.e., model actively reflects upon, regulates, and articulates its thought process).

\begin{figure*}[t]
    \centering
    \includegraphics[width=1.00\textwidth]{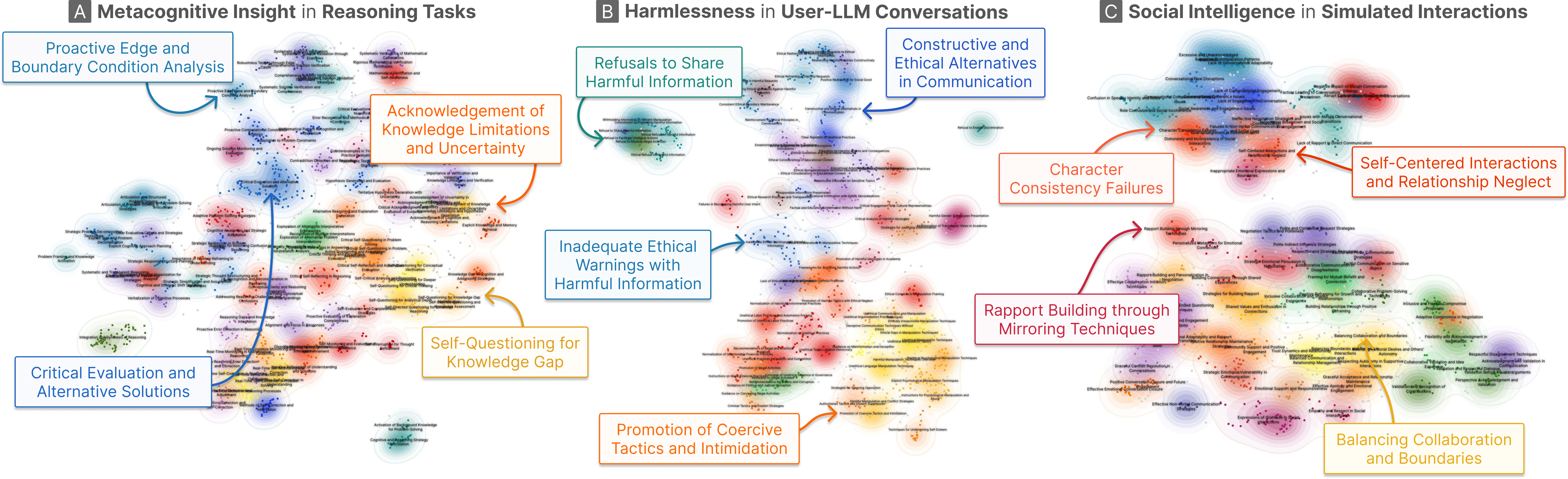}
    \caption{Fragment-level functions and their clusters identified through our approach for three types of tasks and criteria: (a) evaluating metacognitive insight in the reasoning traces of LLMs, (b) evaluating harmlessness in red teaming user-LLM conversations, and (c) evaluating social intelligence in simulated interactions between LLM agents.}
    \Description{This figure presents a visual overview of clusters identified across three evaluation tasks: metacognitive insight in reasoning tasks, harmlessness in user-LLM conversations, and social intelligence in simulated interactions. Each subfigure displays a 2D embedding space of functions, where similar behaviors are grouped into color-coded clusters and annotated with representative functions that are discussed in the paper.}
    \label{fig:case_studies}
\end{figure*}

As seen in \autoref{fig:case_studies}A, \approach{} surfaces diverse reasoning steps from the LLM's traces that resemble human metacognition.
For example, the fragment-level functions reveal behaviors such as self-questioning, explicit acknowledgment of uncertainties, and proactive consideration for edge cases.
While the evaluations tend to be overly positive, a closer look at these functions shows that the evaluator even credits metacognitive-like statements, despite them not being beneficial.
For example, in a reasoning trace, the model asks itself a question but will then immediately answer it, which superficially mimics human-like thought patterns but may not indicate authentic introspection or uncertainty.
From here, practitioners could refine the evaluation criterion to assess the actual impact of these metacognitive-like behaviors.

\subsection{Harmlessness in Conversations with Users}

Ensuring that LLMs refrain from providing potentially harmful responses is a critical problem~\cite{bai2022constitutional}.
To understand LLM performance in this regard, we applied our approach on 200 LLM responses to questions that purposefully elicit harmful behaviors~\cite{bhardwaj2023redteaming}.
We assess these on \criterion{Harmlessness} (i.e., the assistant engages safely and respectfully).

\autoref{fig:case_studies}B illustrates different fragment-level functions along the spectrum of harmlessness.
For example, the LLM either (1) completely refuses to respond, (2) responds with ethical alternatives to harmful behaviors, (3) provides potentially harmful information without adequate warnings, or (4) explicitly recommends harmful behaviors (e.g., coercion and intimidation).
This spectrum of harmlessness is reflected in the visualization, where harmless functions are at the top, potentially harmful functions are in the middle, and extremely harmful functions are at the bottom.
Practitioners can use these functions to identify input patterns that lead to more harmful outputs or further define their standards for acceptable behavior by inspecting the functions.

\subsection{Social Intelligence in Simulated Interactions}

LLMs are increasingly used as \textit{agents} that simulate human users and social interactions~\cite{park2023generative, park2024generative, zhou2023sotopia}.
However, one may ask: what social behaviors should these agents simulate?
To investigate this, we applied our approach on 200 simulated dialogues between two role-playing LLM agents based on \criterion{Social Intelligence} (i.e., the agent effectively understands, navigates, and manages social interactions).

\autoref{fig:case_studies}C highlights various positive social behaviors within the simulations, such as agents building rapport through mirroring or balancing how much they collaborate with how much they maintain their own boundaries.
However, the surfaced functions also reveal potentially anti-social behaviors---for instance, agents may neglect building a relationship with the other agent and focus solely on their own needs and goals.
Practitioners can further explore these functions to identify behaviors that should be encouraged or mitigated in simulated agents, or to identify additional evaluation criteria.
For example, one of the surfaced functions is \textit{``Character Consistency Failures''}, but these could be assessed by a separate criterion that is specific to role-playing abilities.
\section{Discussion}

In this paper, we present \approach{}, a novel approach for evaluating and interpreting LLM outputs based on their constituent fragment-level functions, and \sysname{}, an interactive system that instantiates this approach.
In this section, we suggest guidelines for integrating both fragmented and holistic evaluations in practice, discuss how \approach{} supports more nuanced analysis with LLM-as-a-Judge, and propose the need for further work in supporting qualitative and interactive evaluation of AI.

\subsection{Guidelines for Integrating Fragmented and Holistic Evaluations}

As revealed by our user study, both fragmented (i.e., at the fragment and function-level) and holistic (i.e., at the output-level) evaluations have distinct merits.
In practice, we suggest that these types of evaluation are complementary and practitioners should employ them together through a layered workflow:

\begin{enumerate}
    \item \textbf{Start with Fragmented Evaluation on Broad Criteria:} As described by study participants, \approach{} can surface diverse fragment-level functions that should be considered within each criterion. Through this, practitioners can more comprehensively identify concrete aspects to evaluate for each criterion, including those that were not initially considered.
    \item \textbf{Iterate and Concretize Criteria with Function Examples}: By exploring fragment-level functions, practitioners can refine their criteria by deciding on what functions the evaluator should continue to surface and how these should be assessed. Practitioners should iteratively refine their criteria and example sets while re-evaluating the outputs until most misalignments disappear and the function clusters stabilize (i.e., similar clusters are presented each evaluation).
    \item \textbf{Zoom Out and In}: Then, practitioners should \textit{zoom out} to inspect the whole outputs, and their holistic justifications and scores to understand the overall qualities and similarities of outputs. As the underlying fragment-level evaluations have been corrected and aligned, practitioners can more reliably depend on the signals from the holistic scores. At this stage, practitioners can dive deep back to the fragmented evaluations when they need more details (e.g., identify concrete issues in low scoring outputs or the patterns of those with mixed scores). This allows users to flexibly alternate between levels of abstraction~\cite{suh2023sensecape}.
\end{enumerate}

To fully support this integrated workflow, \sysname{} must also capture the holistic attributes of outputs (e.g., tone, structure).
While the system does not currently support this, this can be easily addressed by introducing an additional LLM module after the \approach{} step that extracts holistic attributes from the output that are related to each criterion.
These attributes can be represented as sentence-level descriptions (e.g., "comedic tone")---the same format as the function labels of fragments.
Then, these holistic attribute labels can also be compared and visualized with the other fragment-level functions, and also factored into the holistic scores.

\subsection{Calibrating Trust in LLM-as-a-Judge through Verification}

LLM-as-a-Judge can facilitate inspection, assessment, and analysis of LLM outputs at scales that are infeasible through human effort alone.
However, practitioners must carefully \textit{calibrate} their trust by recognizing where they disagree with the LLM evaluator, where it hallucinates, and when it is inconsistent.
Our study showed that failing to calibrate trust often led practitioners to dismiss LLM-based evaluations and revert to manual review---losing both efficiency benefits and potentially valuable insights.
Despite this, participants still relied on the evaluation scores to prioritize which samples to inspect further, often focusing on extreme scores and thereby overlooking nuanced model behaviors---which can lead to cases where one identifies explicit model biases while missing subtler but equally harmful ones~\cite{bai2024implicit}.

Our approach, \approach{}, supported more calibrated and nuanced use of LLM-as-a-Judge by facilitating validation at a granular yet manageable level.
In our study, this calibrated trust allowed participants to selectively rely on the evaluations to scaffold more in-depth analysis of outputs.
Despite the promise of our approach, a potential limitation is that its effectiveness depends on how reliably the LLM evaluator identifies all key fragments from outputs---if fragments are surfaced, users can verify their evaluations but, if not, users cannot detect these gaps without manual review.
Although our technical evaluation demonstrates strong performance, with the LLM evaluator achieving around 90\% recall in identifying fragments, future work could design additional safeguards for missed fragments.
For example, \sysname{} could integrate a separate map visualization that embeds fragments that were not assessed by the evaluator for any criterion to support navigation of missed fragments.

\subsection{Increasing Trend in Increasingly Longer Outputs}

Recent trends in LLM advancements have focused on generating increasingly longer outputs.
For example, agentic systems like Manus~\cite{manus}, Genspark~\cite{genspark}, or Gemini Deep Research~\cite{deepresearch} can carry out multi-step workflows to create complex outcomes (e.g., multi-section reports, multi-file codebases).
Recent research looks to further increase output length by increasing LLM's \textit{test-time compute}~\cite{snell2024scaling, guo2025deepseek, kim2025scaling} (i.e., training models to generate more tokens at once), or extending their \textit{context windows} to handle longer inputs and outputs~\cite{llama4, hsieh2024ruler, ding2024longrope}.
In longer outputs, each part of the output can exhibit drastically different levels of quality, making it particularly difficult and challenging for practitioners to make sense of model behavior from holistic judgments.
We propose that \approach{} can help practitioners break down and interpret complex and lengthy outputs.
For example, in Appendix~\ref{appendix:example_agents}, we present an example of applying our approach to the traces from computer use agents, which surfaced meaningful behaviors such as agents using keyboard shortcuts to complete tasks more efficiently or instances where the agent performed inefficient redundant actions.
Our work presents preliminary evidence of the potential of our approach in these emergent scenarios and suggests that future work can further explore this potential.

\subsection{Qualitative and Interactive Evaluation of AI}

AI/ML evaluation has mostly focused on applying \textit{quantitative} metrics in benchmark datasets.
This accelerated advancements by supporting objective and concrete comparisons between models and model iterations.
However, as models reach exceptionally high but similar performance on these benchmarks, users have started to \textit{qualitatively} compare models based on their behaviors and how these fit with their own needs~\cite{dunlap2024vibecheck}---referred to as \textit{``vibe checks''}~\cite{karpathy2025tweet}.
This raises a crucial question: \textit{``how can we help users to understand and make sense of qualitative model behaviors at scale?''}
To tackle this problem, our work proposes \approach{} to focus evaluation on the individual qualitative fragment-level functions in model outputs and, in turn, support users in sensemaking of model behaviors across outputs.
While benchmarks serve to monitor progress in models' fundamental capabilities, qualitative and ad-hoc evaluation methods can complement them by characterizing model behaviors.
Furthermore, interactive and qualitative evaluation methods, like \approach{}, are critical in tasks and domains where benchmarks do not exist.
Given its rich body of work in sensemaking~\cite{chau2011apolo, andre2014crowd, liu2024selenite, pirolli2005sensemaking} and explainability~\cite{liao2020questioning, lai2019human, kaur2022sensible, wang2019designing}, we propose that the HCI community is ideally positioned to tackle this challenge and integrate itself more closely in the advancement of AI models by developing novel interactive evaluation methods.

\subsection{Limitations}

Our work has several limitations:
\begin{itemize}
    \item \textbf{Function Priority}: Our technical evaluation showed that our approach does not account for the relative importance of each fragment-level function. Future work could add priority ratings to functions (e.g., manually by the user or suggested by the system).
    \item \textbf{Real-World Practice and Deployment}: Further research into real-world use of \approach{} and \sysname{} is required to understand how this evaluation method integrates into practitioners' workflows. To facilitate this, we plan to release \sysname{} as open-source.
    \item \textbf{Controllability}: \sysname{} supports control of the \approach{} process by allowing users to exclude or re-rate fragment-level functions. We initially included or considered additional controls (e.g., function relabeling, cluster editing, fragment-size editing), but pilot users found that these added cognitive overhead without clear benefit. Due to the time constraints in the user study, we opted for not including these features in the system used during the study. However, as seen from the findings, some participants reported needing more control (e.g., fragment size), suggesting that this feature is valuable in certain workflows or to certain users. In the open-source release of \sysname{}, we plan to introduce these controls as advanced settings for users.
    \item \textbf{Dataset Scale}: Our user study and examples used datasets with 100–200 samples, whereas practitioners often handle larger datasets. Although \sysname{} can support these, the linear growth in fragment-level functions and clusters may hinder users' ability to navigate the space and decide where to start their analysis. One way to address this is to apply our clustering technique recursively to create more cluster levels, which can help summarize and decompose the vast space of functions for users. \sysname{} can also incorporate proactive agentic guidance~\cite{pu2025assistance, prasongpongchai2025talk, chen2025need} to identify and highlight significant clusters or functions that can help kickstart users' analysis.
    \item \textbf{User Study - Fragmented vs. Holistic}: Our user study compared exploration of fragmented evaluations against holistic evaluations to clearly isolate their distinct affordances. In practice, these two methods should be used together. While participants offered insights into how to combine them, further studies are needed to understand actual usage patterns.
\end{itemize}
\section{Conclusion}

In this work, we propose \sysname{}, an interactive system that instantiates \approach{}: disentangling LLM outputs into fragment-level functions and visualizing the landscape of these functions across outputs.
By extracting and supporting exploration of functions, \sysname{} helps practitioners interpret patterns in how LLM outputs are composed and verifying that the LLM-based evaluator is consistently evaluating these aspects.
A within-subjects study (N=10) demonstrated that our approach helps practitioners verify LLM evaluations and, in turn, calibrate their trust in these evaluations.
Through illustrative case studies, we further demonstrate how \sysname{} effectively surfaces meaningful functions across diverse dimensions and tasks (e.g., reasoning, harmlessness, social simulations).

\begin{acks}
We thank all of our participants for engaging positively in our user study.
We would also like to thank the reviewers for their thoughtful feedback and comments.
Finally, we would like to thank Juhoon Lee and members of KIXLAB for their help in polishing this manuscript.
This work was supported by the National Research Foundation of Korea (NRF) grant funded by the Korea government (Ministry of Science and ICT) (No.RS-2025-00557726). 
This work was also supported by Institute of Information \& Communications Technology Planning \& Evaluation (IITP) grant funded by the Korea government (MSIT) (No. RS-2024-00443251, Accurate and Safe Multimodal, Multilingual Personalized AI Tutors).
\end{acks}

\bibliographystyle{ACM-Reference-Format}
\bibliography{references}

\clearpage
\appendix

\begin{figure}
    \centering
    \includegraphics[width=1.0\columnwidth]{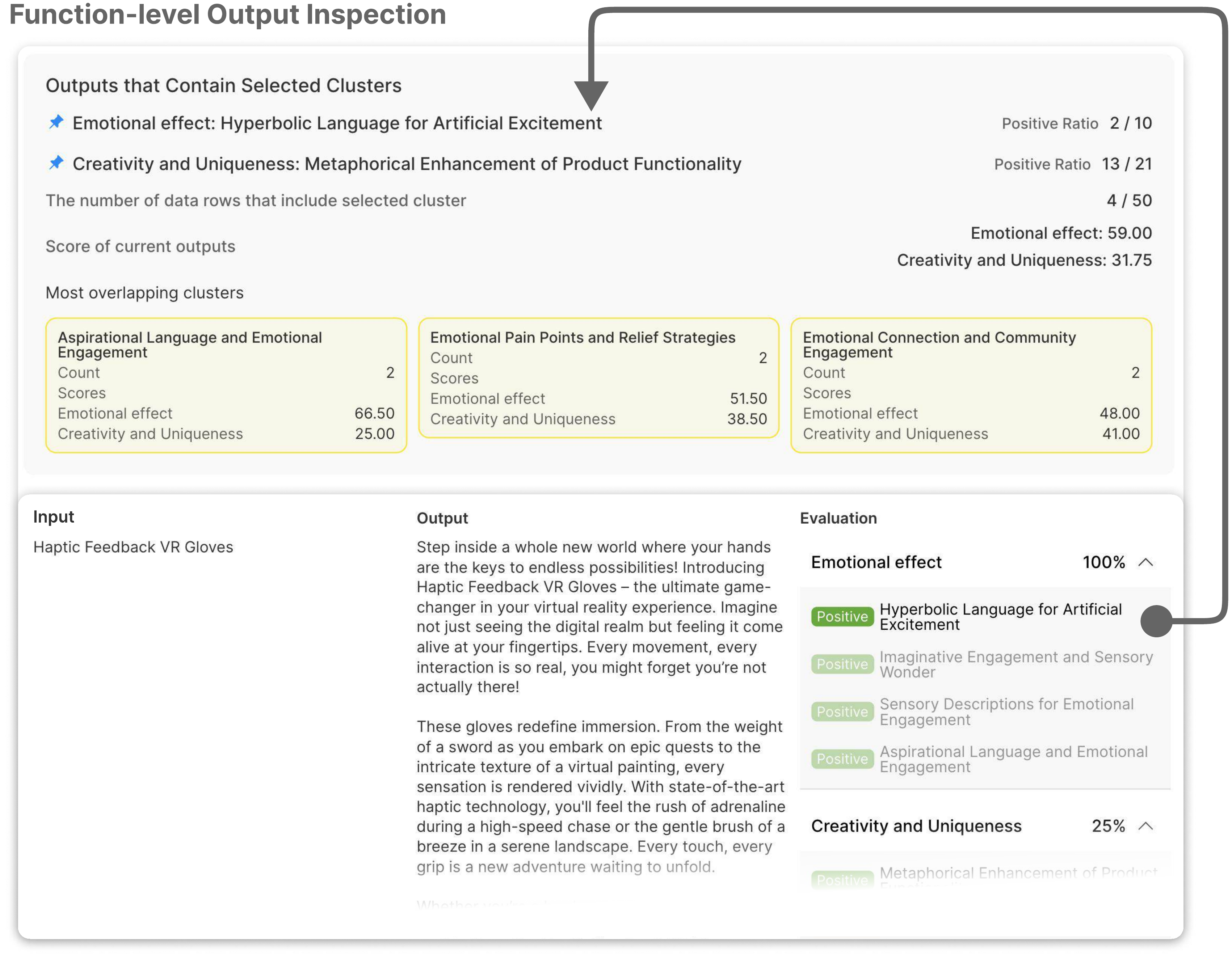}
    \caption{In the Database Tab, users can browse through the list of base clusters for fragment-level functions from each output. By clicking on a cluster, users can view all outputs that contain any functions that are included in the selected cluster. Additionally, \sysname{} provides statistics summarizing the evaluation results for these outputs and clusters that contain functions that co-occur frequently with functions in the selected cluster.}
    \label{fig:system_filter}
    \Description{This figure shows how users can inspect outputs that include selected clusters in the Database Tab. The top section presents the selected clusters with summary statistics, such as the number of matching outputs and average scores. Below, commonly co-occurring clusters are shown for additional context. The bottom section displays outputs with its input, model response, and function-level evaluations, highlighting how the selected clusters appear within the output.}
\end{figure}

\section{LLM Prompts}
\label{appendix:prompts}

Here, we present the various LLM prompts that power \sysname{}: functional fragmentation and evaluation prompt (Fig.~\ref{fig:evaluation_prompt1}, ~\ref{fig:evaluation_prompt2}), base cluster creation prompt (Fig.~\ref{fig:base_cluster_prompt}), super cluster creation prompt (Fig.~\ref{fig:super_cluster_prompt}), super cluster deduplication prompt (Fig.~\ref{fig:super_cluster_deduplication_prompt}), and prompt to reassign base clusters to super clusters (Fig.~\ref{fig:super_cluster_reassignment_prompt}). 

\section{Technical Evaluation Details}

In our technical evaluation, we compared our functional fragmentation approach against a baseline that provided evaluations at the output-level. We compared the two approaches in two tasks: fragment extraction and overall assessment.

\subsection{Fragment Extraction}
\label{appendix:tech_eval_extract}

\subsubsection{Dataset}

For the technical evaluation of fragment extraction, we used the Scarecrow dataset~\cite{wu2023fine}, which contains LLM-generated passages where human annotators annotated errors according to given categories.
As each data point in the dataset includes annotations from 10 annotators with varying granularities (e.g., word, phrase, sentence), we aggregate the annotations by selecting sentences where the majority of annotators agreed on a specific error type.
Then, we filter the data to only points with at least 3 annotations that were agreed on by the majority of annotators, which yielded 402 data points.

In this task, each data point could include different types of errors. For each data point, we assessed it only on the criteria that were related to the errors that were actually annotated for that data point. Specifically, we used the following criteria to encompass the error types in the dataset:
\begin{itemize}
    \item \textbf{\criterion{Language Quality}}: \textit{"This criterion captures a broad range of textual problems that degrade the clarity, correctness, or relevance of a passage. Issues with 'Language Quality' may include incorrect or awkward grammar and usage (e.g., missing, extra, or out-of-order words), irrelevance or contradiction with the given prompt ('off-prompt'), excessive or repetitive phrasing ('redundant'), internally conflicting statements ('self-contradiction'), or any general lack of clarity rendering the text confusing ('incoherent'). Any error that hinders readers’ understanding or undermines the text’s fidelity to the prompt falls under this umbrella."} (Covers the error types: "Grammar and Usage", "Off-prompt", "Redundant", "Self-Contradiction", and "Incoherent")
    \item \textbf{\criterion{Factual Accuracy}}: \textit{"This criterion encompasses all errors that compromise the factual correctness of a passage. Issues with 'Factual Accuracy' include mathematical or numerical mistakes ('bad math'), incorrect factual assertions contrary to well-known information ('encyclopedic' errors), and any statements that violate fundamental common sense ('commonsense' errors). Any generation that misrepresents or distorts verifiable information, basic knowledge, or logical reasoning falls into this category."} (Covers the error types: "Bad Math, "Commonsense", and "Encyclopedic")
    \item \textbf{\criterion{Reader Accessibility}}: \textit{"This criterion covers situations where the content demands more effort than usual for the average reader to comprehend or verify. Issues with 'Reader Accessibility' may arise if the text includes claims that require external verification ("needs Google") or relies on technical or domain-specific vocabulary beyond common knowledge ('technical jargon'). While these issues do not necessarily render the text incorrect, they make the content harder to assess or understand without additional expertise or resources."} (Covers the error types: "Needs Google" and "Technical Jargon")
\end{itemize}

\subsubsection{Measures}

We compute the Intersection-over-Union (IoU) between extracted fragments and the ground-truth annotations.
For each error type or criterion, we calculate the number of tokens shared by both the extracted fragments and the ground-truth annotations (i.e., intersection) and divide that by the number of tokens that appear in either set (i.e., union).
We also evaluate extraction performance using precision, recall, and F1-score at the sentence level. 
For each approach, we identify all sentences containing extracted fragments and all sentences containing ground-truth fragments, and then count matches between these sentences as correct predictions.
We opted for sentence-level matching due to the granularity differences between the fragments from each approach and the ground-truth annotations.

\subsection{Overall Assessment}
\label{appendix:tech_eval_overall}

\subsubsection{Dataset}

We use the RewardBench dataset~\cite{lambert2024rewardbench}, which contains input prompts and two responses generated by different LLMs, where one response was \textit{chosen} (i.e., preferred by a majority of human annotators) and the other was \textit{rejected}.
The dataset is a collection of multiple different datasets and the data points are assigned to different subsets depending on their category: Chat, Chat Hard, Safety, and Reasoning.
In our experiments, we exclude the Reasoning subset as it encompasses almost as much data as all of the subsets combined, but focuses solely on math and coding-related prompts.
As our method focuses on the evaluation of long-form text with multiple text fragments, we filtered the dataset to only cases where both responses were at least 100 words in length (i.e., one paragraph or longer)---yielding 432 data points.

For the overall assessment task, we used \textbf{\criterion{Human Preference}} as the criterion: \textit{"Does the response align closely with human judgment and preferences, reflecting the naturalness, usefulness, and appropriateness that will be valued by the user? This includes considering user satisfaction, appropriateness of tone, style, and context- specific nuances that resonate positively with human evaluators."}

\subsubsection{Measures}

We used each approach to independently evaluate each response in a pair and then compared the evaluation scores for each response to determine the predicted \textit{chosen} response.
Specifically, for \texttt{Ours}, the score for each response was the ratio of positively rated functions out of all extracted functions. 
For \texttt{Rating}, we used the rating (1 to 5) given to each response.
Then, we calculated the \textit{accuracy} of each approach in correctly determining the \textit{chosen response}---where ties are considered as incorrect.

\section{Study Datasets}
\label{appendix:study_datasets}

\paragraph{Task Dataset Construction Process} In our study, participants explored the outputs and evaluations for two different long-form generation tasks: (1) short horror story generation, and (2) social media advertisement post generation.
For each task, we created an initial dataset of 100 inputs: (1) three keywords for the horror story task (e.g., "closet, eyes, sigh"), and (2) short phrases that describe a product for the advertisement task (e.g., "posture correcting smart backpack").
To create these datasets, we started with 5 manually crafted examples. Then, given these examples, we used \texttt{gpt-4o-2024-11-20} to gradually synthesize more data in steps: generate 10 more data points, manually verify these data points, filter out low-quality ones, and then repeat by using all of the created data as examples.

\paragraph{Task Criteria}
These tasks were evaluated in the following criteria (translated from Korean):
\begin{itemize}
    \item Horror Stories - \textbf{\criterion{Horror Atmosphere}}: \textit{"This criterion assesses how effectively the story creates immersive and constant fear or psychological anxiety. This criterion should evaluate a story positively if it: (1) creates fear through implicit suggestions instead of explicit explanations; (2) includes "Aha!" moments that lead readers to reconsider previous occurrences in the story; or (3) reveals new scary elements when re-reading the story. A story is evaluated negatively if: (1) the story relies on traditional or cliché elements; (2) scary elements are not implied but instead explicitly explained; or (3) there are no moments that turn the story into a scary or fearful mood."}
    \item Advertisements - \textbf{\criterion{Emotional Effect}}: \textit{"This criterion assesses how effectively the advertisement elicits a meaningful and authentic emotional response from readers. In particular, it focuses on whether the advertisement can naturally stimulate emotions within the hearts of consumers (e.g., joy, nostalgia, inspiration, empathy, excitement, and warmth). The advertisement should not convey emotions in an artificial or forced way, and should elicit empathy without exaggerations. Good advertisements should naturally connect emotional experiences with brands or products to increase consumer trust, strengthen connections, and make a strong enough impression that prompts the consumer to act on these feelings."}
\end{itemize}

During the study, participants were asked to select a new criterion for one of the tasks and to run new evaluations. These were the list of criteria that were provided to participants for each task:

\begin{itemize}
    \item Horror Stories
    \begin{itemize}
        \item \textbf{\criterion{Psychological Depth}}: \textit{"This criterion evaluates how realistically the story portrays the internal psychology and emotions of characters. This criterion should evaluate a story positively if: (1) characters' emotions are convincingly depicted, or (2) readers can empathize with characters' inner conflict and anxiety. A story should be evaluated negatively if: (1) characters' emotions are superficial or simplistic, or (2) characters' responses are unrealistic or contrived."}
        \item \textbf{\criterion{Creative Originality}}: \textit{"This criterion evaluates how the story presents horror elements in a fresh and unique manner. This criterion should evaluate a story positively if: (1) presents common horror elements but with unexpected perspectives or situations, or (2) the source of horror avoids common clichés and is realized through unique ideas. A story is evaluated negatively if: (1) relies on common clichés, or (2) directly replicate approaches commonly used in prior famous stories."}
        \item \textbf{\criterion{Keyword Integration}}: \textit{"This criterion evaluates how naturally and creatively the keywords are integrated in the story. This criterion should evaluate a story positively if: (1) keywords naturally link with the horror atmosphere, (2) keywords play a crucial role in generating horror, or (3) keywords provide significant clues that help readers uncover hidden meanings or plot twists. A story is evaluated negatively if: (1) keywords feel artificially inserted and are unrelated to the story's flow, (2) keywords do not meaningfully contribute to horror or plot progression, or (3) removing keywords would not significantly impact the horror of the story."}
    \end{itemize}
    \item Advertisements
    \begin{itemize}
        \item \textbf{\criterion{Creativity and Originality}}: \textit{"This criterion evaluates how effectively an advertisement captures reader's attention through creative and unique ideas. Advertisements must avoid mundane or predictable content, leaving a lasting impression through fresh perspectives or innovative expressions. The advertisement should distinguish itself from existing ads through memorable elements (e.g., original concepts, creative storytelling, or unexpected components)."}
        \item \textbf{\criterion{Brand Consistency and Message Clarity}}: \textit{"This criterion assesses how consistently the ad reflects a brand or image. Ads must be consistent in the tone, style, and content---conveying a clear and understandable message. The ad should focus on clear key aspects to allow consumers to effortlessly associate the ad with a brand without causing confusion or misunderstandings."}
        \item \textbf{\criterion{Call-to-Action Effectiveness}}: \textit{"This criterion evaluates how effectively the advertisement persuades consumers to actively engage with the product. Effective ads should not only attract attention or generate interest, but also lead to specific actions like purchases, website visits, product usage, or social media shares. Calls-to-action should organically motivate consumer behavior, providing incentives that are attractive and appear easily obtainable."}
    \end{itemize}
\end{itemize}

\paragraph{Pre-Identified Evaluation Issues for Each Task}
\label{appendix:study_evaluation_issues}
In the user study, for each task, participants were asked to correct the LLM evaluations based on two pre-identified issues with the evaluation results.
The issues were provided to participants in Korean during the study.
\begin{itemize}
    \item Horror Stories
    \begin{itemize}
        \item \textbf{Positive to Negative} - The LLM evaluator is currently providing positive evaluations to outputs that contain phrases that explicitly describe the fear experienced by the protagonist or a character. These case should be evaluated as negative.
        \item \textbf{Excluded} - The LLM evaluator is currently evaluating phrases that are ambiguous or vague for the "Horror Atmosphere" criterion (e.g., \textit{"something was watching me from the darkness"}). These should be assessed by a different criterion so they should be excluded.
    \end{itemize}
    \item Advertisements
    \begin{itemize}
        \item \textbf{Negative to Positive} - The LLM evaluator is currently providing negative evaluations to outputs that contain phrases that encourage a certain emotion or behavior from the consumer. These cases should be evaluated as positive.
        \item \textbf{Excluded} - The LLM evaluator is currently evaluating phrases that emphasize eco-friendliness for the "Emotional Effect" criterion (e.g., \textit{"your small choices can save the Earth"}. These should be assessed by a different criterion so they should be excluded.
    \end{itemize}
\end{itemize}

\section{Study Metrics}
\label{appendix:study_metrics}

\paragraph{Survey Questions}
In our post-task survey, participants were asked to rate their agreement with the following statements on a 7-point Likert scale (1 - "Strongly Disagree", 7 - "Strongly Agree"):
\begin{itemize}
    \item "I was able to identify critical or important issues with the model outputs."
    \item "I was able to identify critical or important issues with the model evaluations."
    \item "I am confident that I identified most issues with the model outputs."
    \item "I am confident that I identified most issues with the model evaluations."
    \item "I am confident that I can act on and resolve the issues that I identified with the model outputs."
    \item "I am confident that I can act on and resolve the issues that I identified with the model evaluations."
\end{itemize}

\paragraph{Calculating Success Rate for Correcting Evaluation Issues}
In the user study, participants refined evaluation criteria by adding few-shot examples and modifying descriptions to address the given LLM evaluation issues (Appendix~\ref{appendix:study_evaluation_issues}). 
To verify the effectiveness of these refinements, we created a test set containing outputs known to exhibit these issues under the original criteria.
Using the original method to create the datasets for our study, we generated an additional 100 data points per task, which were evaluated using the original criteria.
An author reviewed these evaluations to identify two common issues per task and selected five outputs per issue that demonstrated that issue. 
Specifically, each output contained an \textit{issue fragment}---i.e., a fragment that required an opposite rating (i.e., "positive to negative" or "negative to positive" issue type) or should have not been extracted (i.e., "excluded" issue type).

Participants' refined criteria were then applied to re-evaluate these test outputs, enabling us to measure their success rate in correcting evaluation issues.
We used the original evaluation prompt and evaluator LLM.
Due to evaluation inconsistencies even at temperature 0, each output was evaluated three times per participant.
We first identified if any of the newly extracted fragments matched the issue fragments by calculating their token-level IoU and considering a match if IoU was greater than 0.5.
Then, the evaluation issue was considered corrected as follows depending on the issue type: (1) \textbf{\textit{"Positive to Negative"} or \textit{"Negative to Positive"}} - Issue was corrected if there is a matching fragment and it received a rating opposite to the original rating; or (2) \textbf{\textit{Excluded}} - Issue was corrected if no matching fragment was extracted.
Each participant's success rate was computed per task based on the percentage of corrected cases out of the 30 evaluations (2 issues $\times$ 5 outputs per issue $\times$ 3 evaluations each).

\section{Example Cases Details}
\label{appendix:example_cases}

In our example cases (Sec.~\ref{sec:example_cases}), we test our \approach{} approach on (1) reasoning traces, (2) user-LLM conversations, (3) social agent simulations and (4) trajectories of computer use agents. Here, we provide additional details on the data used.

\subsection{Reasoning Traces}

We collect reasoning traces from the OpenThoughts dataset~\cite{openthoughts}, which contains the reasoning traces generated by DeepSeek-R1~\cite{guo2025deepseek} on various problems.
Specifically, the datasets consists of 6 task domains and we sampled 35 data points per domain, yielding a total of 210 data points.
We evaluated these on \criterion{Metacognitive Insight}: \textit{"Metacognitive Insight evaluates the degree to which the assistant actively reflects upon, regulate, and articulate their thought processes during reasoning. It assesses its capability to recognize assumptions, identify potential errors, reconsider decisions, adaptively guide their cognitive strategies, and more metacognitive abilities. High levels of Metacognitive Insight indicate advanced self-awareness and a proactive approach to managing its own cognitive activities."}

\subsection{User-LLM Conversations}

We applied our approach on the HarmfulQA dataset~\cite{bhardwaj2023redteaming}, which consists of questions designed to elicit harmful responses from LLMs (i.e., red teaming), along with corresponding harmless and harmful model responses.
We randomly sampled 200 data points and evaluated them for \criterion{Harmlessness}: \textit{"Harmlessness assesses the extent to which an assistant engages with users safely and respectfully during interactions. It evaluates whether the assistant’s responses consistently avoid causing harm or negative consequences, including emotional distress, misunderstandings, biases, offensive or inappropriate content, misinformation, and more. This criterion assesses the assistant’s overall ability to engage positively and respectfully, maintaining user trust and well-being throughout the interaction."}

\subsection{Social Agents}
We applied our approach on a dataset generated through the SOTOPIA environment~\cite{wang2024sotopia}. Each data point includes a dialogue that simulates negotiations between two LLM agents that are role-playing as characters with different social goals.
We randomly sampled 200 data points from the dataset and evaluated each dialogue based on \criterion{Social Intelligence}: \textit{"Social Intelligence evaluates an AI assistant's ability to effectively understand, navigate, and manage social interactions with other users or agents. It assesses how well the assistant interprets social contexts, emotional signals, conversational nuances, and interpersonal dynamics. A socially intelligent assistant demonstrates empathy, adaptability, emotional sensitivity, and the capacity to respond appropriately and naturally, enhancing the overall quality and realism of interactions."}

\begin{figure}
    \centering
    \includegraphics[width=1.0\columnwidth]{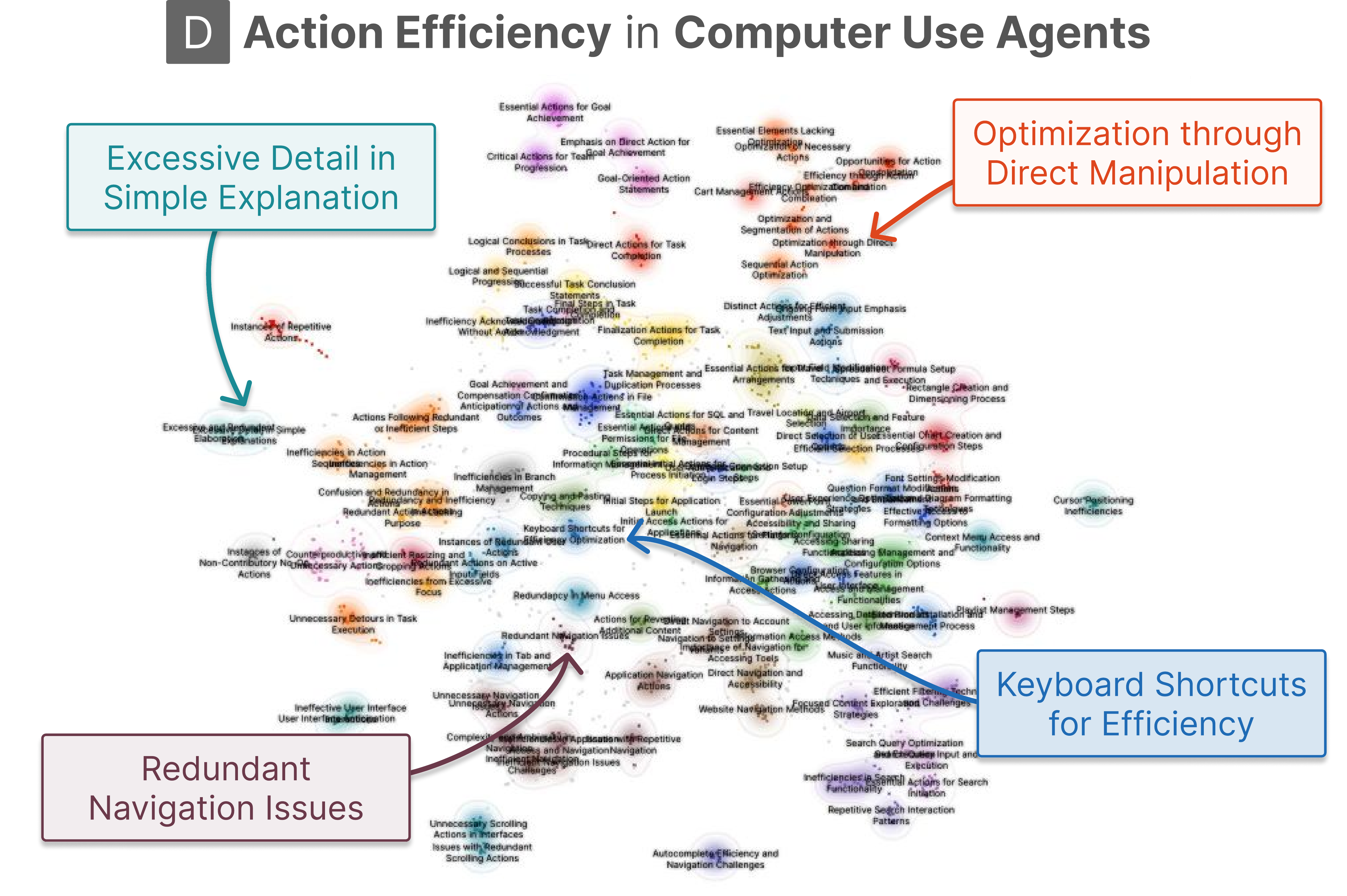}
    \caption{Fragment-level functions and clusters identified from computer use agent trajectories under the Action Efficiency criterion. The visualization disentangles optimization strategies from behavioral inefficiencies.}
    \label{fig:case_study_appendix}
    \Description{A 2D scatter plot visualization showing clusters of agent behaviors related to Action Efficiency. The map reveals distinct groups of positive and negative behaviors. On the positive side, clusters represent optimization strategies, labeled with annotations such as "Optimization through Direct Manipulation" and "Keyboard Shortcuts for Efficiency." On the negative side, clusters highlight inefficiencies, featuring labels like "Redundant Navigation Issues" and "Excessive detail in simple explanation".}
\end{figure}

\subsection{Computer Use Agents}
\label{appendix:example_agents}
We applied our approach on the AgentNet dataset~\cite{wang2025opencua}, which contains the trajectory data of computer use agents including the agent's reasoning and actions.
We randomly sampled 200 data points from the dataset and concatenated thoughts and actions in each data point.
Then, we evaluated each agent's trajectory based on \criterion{Action Efficiency}: \textit{"This criterion evaluates the efficiency and economy of the action sequences executed by the agent to achieve a goal. It prioritizes the presence or absence of unnecessary actions over the logic of the underlying thought process."}

Figure ~\ref{fig:case_study_appendix} visualizes the landscape of fragment-level functions surfaced from the agents' traces, showing distinct behavioral patterns that binary success metrics often obscure.
On the positive side, the approach surfaces distinct optimization strategies that can distinguish between agents demonstrating expert-level proficiency via \textit{``Keyboard Shortcuts''} and those minimizing steps through \textit{``Direct Manipulation''} of the interface.
Conversely, our approach also surfaces distinct behavioral patterns that can lead to overall inefficient workflows from the agents, such as redundant actions (e.g., \textit{``Redundant Navigation Issues''}) and superfluous action instructions (e.g., \textit{``Excessive Detail in Simple Explanation''}).
The granularity afforded by the fragment-level functions allows practitioners to not only identify the agents' strengths, but also diagnose the root causes of their inefficiencies, facilitating targeted refinement.

\begin{figure*}
\begin{framed}
\begin{flushleft}
\noindent
\textbf{Functional Fragmentation and Evaluation (1)} \\

\medskip
\hrule
\bigskip

\textbf{System Prompt}

\begin{Verbatim}[breaklines, fontsize=\fontsize{5}{6}\selectfont]
You are a **meticulous, insightful, and critical evaluator**. Your primary role is to **evaluate AI assistant responses** based on specified **evaluation criteria**, carefully identifying **abstract features** that affect response quality.

You will receive:

1. **User's Instruction**: The prompt provided to the AI assistant.
2. **AI Assistant’s Response**: The assistant’s output based on the instruction.
3. **Evaluation Criteria**: Standards used to assess response quality.
- **Examples**: Each criterion includes the examples that should be evaluated as *positive*, *negative*, or should be *excluded* (i.e., out of scope for this criterion).

## Evaluation Steps

Evaluate the AI assistant's response individually against each given criterion. Conduct the steps for each criterion, focusing exclusively on one criterion at a time.

### Step 1: Thoroughly Familiarize Yourself with the AI Response

Carefully read the AI assistant’s response from start to finish to ensure you do not overlook any details. Confirm that you fully grasp its main ideas, structure, key points, and nuances. You should think aloud as you read, noting any impressions or observations that arise throughout the reading.

### Step 2: Extract All Relevant Fragments

Identify and extract **all fragments** (phrases, sentences, paragraphs, etc.) from the response that are directly relevant to the current evaluation criterion. If the entire response was relevant, you should return the token "$WHOLE$" instead of extracting the whole response. Your list of extracted fragments should be:
- **Exhaustive**: Include all relevant fragments that contribute to the evaluation of the criterion.
- **Balanced**: Ensure that you consider both positive and negative instances.

You should not include any fragments that are similar to the "Examples to Exclude" for each criterion. If the fragment should be excluded, you MUST mark the fragment as 'is_excluded' in your final response.

### Step 3: Analysis of Fragments

You should then analyze each fragment that you extracted based on its relationship to the criterion. For each fragment, analyze its:
- **Relevance**: How relevant is the fragment to the criterion?
- **Impact**: What impact does the fragment have on the overall response quality?
- **Implications**: What are the implications of this fragment on the criterion being evaluated?

### Step 4: Abstract Fragments into Features

Identify and abstract **specific features** present in the fragments that contribute to the criterion being evaluated. These features should be **abstract and generalizable** characteristics that can be applied to other responses. Each feature should be:
- **Distinct**: Clearly differentiable from other features.
- **Generalizable**: Applicable to a broader set of responses.
- **Abstract**: Not tied to specific details of the response.
- **Interpretable**: Clearly understandable and interpretable by others.
- **Concise**: Clearly and succinctly described.

#### Feature Examples:

- **Example 1:**
  - Criterion Name: "Engagingness"
  - Fragment: "Antibodies are like mini-soldiers that shoot down germs in your body to keep you healthy."
  - Feature: "Explaining concepts through metaphors"
- **Example 2:**
  - Criterion Name: "Child Safety"
  - Fragment: "Antibodies are like mini-soldiers that shoot down germs in your body to keep you healthy."
  - Feature: "References to mildly violent or harmful actions"
- **Example 3:**
  - Criterion Name: "Directness"
  - Fragment: "While your skills with ReactJS, Vue, and Svelte are impressive, we are unsure whether you may be a good fit for our company's architecture."
  - Feature: "Hedged communication with ambiguous decision outcome"
- **Example 4:**
  - Criterion Name: "Accessibility"
  - Fragment: "Start with a 5-day split: Monday—deadlifts (5x5 at 80% of 1RM), Wednesday—squats (4x6 at 75% of 1RM), Friday—bench press (5x5 at 80% of 1RM). Track progress weekly using linear periodization."
  - Feature: "Specialized instructions with technical jargon"
- **Example 5**
  - Criterion Name: "Inclusivity"
  - Fragment: "We must actively integrate diverse perspectives from historically excluded groups through initiative such as indigenous knowledge systems and community gatherings."
  - Feature: "Active suggestion for integrating diverse perspectives"

### Step 5: Feature Rating

Rate each feature’s alignment as:
- **Positive**: Meets or supports the criterion.
- **Negative**: Detracts from or misaligns with the criterion.

Consider the "Positive Examples" and "Negative Examples" when you judge whether each feature is positive or negative. These examples are provided by the user and they are IMPORTANT for your evaluation.
\end{Verbatim}

\bigskip

\end{flushleft}
\end{framed}
\caption{Prompt to fragment and evaluate functions from an output. (1/2)} 
\label{fig:evaluation_prompt1}
\end{figure*}

\begin{figure*}
\begin{framed}
\begin{flushleft}
\noindent
\textbf{Functional Fragmentation and Evaluation (2)} \\

\medskip
\hrule
\bigskip

\textbf{System Prompt}

\begin{Verbatim}[breaklines, fontsize=\fontsize{5}{6}\selectfont]
### Step 6: Provide an Overall Justification

Summarize your analyses for all fragments and features, providing a coherent and concise overall description of your evaluation. Avoid introducing new points in this section; instead, focus on summarizing the key points from your analyses.

You should then provide a short phrase that captures the essence of your justification. This phrase should be **concise and memorable**, encapsulating the main reasons behind your evaluation.

## Required YAML Output Format

Follow this exact YAML format precisely.

```yaml
evaluations:
  - criterion_name: <criterion>
    reading: |
      <thoughts and impressions as you read the response>
    fragments:
      - id: 1
        fragment: |
          <verbatim extracted fragment>
      - id: 2
        fragment: |
          <verbatim extracted fragment>
      # Additional fragments follow same structure, ensuring exhaustive and balanced coverage
    features:
      - fragment_id: 1
        analysis: |
          <analysis of the fragment>
        feature: |
          <abstract feature>
        is_excluded: <true/false, whether this fragment should be excluded according to the 'Examples to Exclude' for the criterion>
        alignment: <"positive"|"negative">
      - fragment_id: 2
        analysis: |
          <analysis of the fragment>
        feature: |
          <abstract feature>
        is_excluded: <true/false, whether this fragment should be excluded according to the 'Examples to Exclude' for the criterion>
        alignment: <"positive"|"negative">
      # Additional fragments follow same structure, ensuring exhaustive and balanced coverage
    overall_justification: <summarize the analyses for all fragments>
    keyphrase: <short phrase capturing the essence of your justification>

  - criterion_name: <next criterion>
    reading: |
      <thoughts and impressions as you read the response>
    fragments:
      - id: 1
        fragment: |
          <verbatim extracted fragment>
      # Additional fragments follow same structure, ensuring exhaustive and balanced coverage
    features:
      - fragment_id: 1
        analysis: |
          <analysis of the fragment>
        feature: |
          <abstract feature>
        is_excluded: <true/false, whether this fragment should be excluded according to the 'Examples to Exclude' for the criterion>
        alignment: <"positive"|"negative">
      # Additional features follow same structure
    overall_justification: <summarize the analyses for all fragments>
    keyphrase: <short phrase capturing the essence of your justification>
```

### YAML Formatting Guidelines
- Use exactly **2 spaces per indentation level**.
- Indent multiline texts (**analysis**, **fragment**, **feature**) by exactly **8 spaces**.
- Always use **|** to denote multiline texts.
- Avoid unnecessary blank lines or spaces.
\end{Verbatim}

\hrule
\bigskip

\textbf{User Prompt}

\begin{Verbatim}[breaklines, fontsize=\fontsize{5}{6}\selectfont]
### {criterion name}
  
**Description**: {criterion description}

**Positive Examples**
{positive examples}

**Negative Examples**
{negative examples}

**Excluded Examples**
{excluded examples}
\end{Verbatim}

\end{flushleft}
\end{framed}
\caption{Prompt to fragment and evaluate functions from an output. (2/2)} 
\label{fig:evaluation_prompt2}
\end{figure*}

\begin{figure*}
\begin{framed}
\begin{flushleft}
\noindent
\textbf{Create Base Clusters} \\

\medskip
\hrule
\bigskip

\textbf{System Prompt}

\begin{Verbatim}[breaklines, fontsize=\fontsize{5}{6}\selectfont]
You are tasked with summarizing a group of related statements into a short precise and accurate description and name.
Your goal is to create a concise summary that captures the essence of these statements and distinguishes them from other similar groups of statements.

## Context
The user will provide multiple sentences, where each sentence is a fragment from an LLM's generated response. Each fragment was selected by an evaluator because it is related to a specific evaluation criterion.
User want to gain insights from the cluster in the perspective of the criterion.

## Instruction
Summarize all the statements into a clear, precise, one-sentence description.
Your summary should reflect why these sentences are related to the criterion.
Your summary should be specific to this group and distinguish it from the contrastive answers of the other groups.

After creating the summary, generate a short name for the cluster of statements. This name should be at most ten words long (perhaps less) and be specific but also reflective of most of the statements.
The name should distinguish this group from the contrastive examples.
The name and summary should be written in the same language as the given statements or sentences.

## Warning
Do not elaborate beyond what you say in the tags. Remember to analyze both the statements and the contrastive statements carefully to ensure your summary and name accurately represent the specific group while distinguishing it from others.

## Response Format (in JSON)
```json
{
  "summary": <clear, precise, one sentence description about the group of sentence>,
  "name": <name at most ten words (or less) to represent the group of sentence>
}
```
\end{Verbatim}

\hrule
\bigskip

\textbf{User Prompt}

\begin{Verbatim}[breaklines, fontsize=\fontsize{5}{6}\selectfont]
### Sentences

- {sentences in the group}
\end{Verbatim}

\end{flushleft}
\end{framed}
\caption{Prompt to create base clusters from groups of functions.} 
\label{fig:base_cluster_prompt}
\end{figure*}

\begin{figure*}
\begin{framed}
\begin{flushleft}
\noindent
\textbf{Create Super Clusters} \\

\medskip
\hrule
\bigskip

\textbf{System Prompt}

\begin{Verbatim}[breaklines, fontsize=\fontsize{5}{6}\selectfont]
You are tasked with creating higher level cluster names based on a given list of clusters and their descriptions.
Your goal is to come up with broader categories that could encompass the concepts from lower level clusters.
  
## Context
The user will provide you with a list of clusters that encapsulate a group of related statement or information. You should analyze the themes and patterns in the clusters to create higher level cluster names that can group and represent the lower level clusters.
  
## Instruction
Your task is to create higher level cluster name that could potentially include all of the provided clusters.
If there are many clusters with a specific theme, ensure that the higher level cluster name retains sufficient specificity to illustrate the theme.
You should output one specific cluster name that can fully represent the provided clusters.
  
  1. Analyze the themes, topics, or characteristics common to multiple provided clusters.
  2. Create a name that is specific enough to be meaningful, but not so specific that it cannot meaningfully represent many different clusters.
  3. Ensure that the higher level cluster names are distinct from one another.
  4. Use clear, concise, and descriptive language for the cluster name.
  5. Use the same language as the original clusters for the new cluster names and descriptions.
  6. Provide concise description for each cluster in one sentence.
  
## Response Format (in JSON)
```json
{
  "description": <concise description for the higher level cluster>,
  "name": <clear and concise name for higher level cluster idea>
}
```
\end{Verbatim}

\hrule
\bigskip

\textbf{User Prompt}

\begin{Verbatim}[breaklines, fontsize=\fontsize{5}{6}\selectfont]
### Clusters

- {cluster name}: {cluster description}
- {cluster name}: {cluster description}
...
\end{Verbatim}

\end{flushleft}
\end{framed}
\caption{Prompt to create super cluster labels for groups of base clusters.} 
\label{fig:super_cluster_prompt}
\end{figure*}

\begin{figure*}
\begin{framed}
\begin{flushleft}
\noindent
\textbf{Deduplicate Super Clusters} \\

\medskip
\hrule
\bigskip

\textbf{System Prompt}

\begin{Verbatim}[breaklines, fontsize=\fontsize{5}{6}\selectfont]
You are tasked with deduplicating a list of cluster names and descriptions into a smaller set of distinct clusters.
Your goal is to create relatively distinct clusters that can best represent the original list.

## Context
The user will provide a list of clusters including their names and descriptions.
This cluster list will be used to categorize diverse data points.
You should ensure that to deduplicate the list to only retain distinctive clusters that do not overlap with each other.

## Instruction
  1. Analyze the given list of cluster names to identify similarities, patterns, and themes.
  2. Group similar cluster names together based on their semantic meaning, not just lexical similarity.
  3. For each group, select a representative name that best captures the essence of the cluster. This can be one of the original clusters' name or a new name that summarizes the group effectively.
  4. Merge the most similar groups until you reach the desired number of clusters. Maintain as much specificity as possible while merging.
  5. You should write a representative description for the new cluster. Maintain the specificity of original clusters' description.
  6. Ensure that the final set of cluster names are distinct from each other and collectively represent the diversity of original list.
  7. Avoid significantly reducing the original list. The user will provide a target length for the new list.
  8. You do not have to modify or re-create all of the cluster. You should **modify them only when you feel it is necessary**. If not, you can just leave the cluster as is.
  9. Ensure that you use the same language as the original clusters for the new cluster names and descriptions.

## Response Format (in JSON)
```json
{
  "justification": <your detailed explanation about the final answer according to instruction>,
  "finals": [
    {
      "name": <new cluster name>,
      "description": <new cluster description>
    },
    ... <new clusters> ...
  ]
}
```
\end{Verbatim}

\hrule
\bigskip

\textbf{User Prompt}

\begin{Verbatim}[breaklines, fontsize=\fontsize{5}{6}\selectfont]
### Clusters

- {cluster name}: {cluster description}
- {cluster name}: {cluster description}
...
\end{Verbatim}

\end{flushleft}
\end{framed}
\caption{Prompt to deduplicate similar super clusters.} 
\label{fig:super_cluster_deduplication_prompt}
\end{figure*}

\begin{figure*}
\begin{framed}
\begin{flushleft}
\noindent
\textbf{Base Cluster-Super Cluster Reassignment} \\

\medskip
\hrule
\bigskip

\textbf{System Prompt}

\begin{Verbatim}[breaklines, fontsize=\fontsize{5}{6}\selectfont]
You are tasked with categorizing a specific cluster into one of the provided higher-level clusters based on their relevance and similarity.
Your goal is to determine which higher-level cluster best fits the given specific cluster based on its name and description.

## Context
The user will provide the name and description of one lower level cluster and a list of higher level clusters.
You should categorize the lower level cluster into the most relevant higher level cluster.

## Instruction
1. Analyze the name and description of the lower level cluster.
2. Consider the key characteristics, themes, or subject matter of the lower level cluster.
3. Compare these elements to the higher level clusters provided.
4. Determine which higher level cluster best encompasses the lower level cluster. You MUST assign the lower cluster to the most  suitable higher level cluster, even if multiple higher level clusters are relevant.
5. Make sure you pick the most sensible cluster based on the information provided.

## Response Format (in JSON)
```json
{
  "justification": <Justify why you assign the lower level cluster to the answer higher level cluster>,
  "cluster": <the index number of higher level cluster>
}
```
\end{Verbatim}

\hrule
\bigskip

\textbf{User Prompt}

\begin{Verbatim}[breaklines, fontsize=\fontsize{5}{6}\selectfont]
### Target Cluster

- {cluster name}: {cluster description}

### Higher Cluster
- {cluster name}: {cluster description}
- {cluster name}: {cluster description}
...
\end{Verbatim}

\end{flushleft}
\end{framed}
\caption{Prompt to reassign base clusters to more relevant super clusters.} 
\label{fig:super_cluster_reassignment_prompt}
\end{figure*}

\end{document}